%% file: nematic_hyper3.tex
\documentclass[aps,pre,preprint,onecolumn,citeautoscript,superscriptaddress,
eqsecnum]{revtex4-1}  
\usepackage{amsmath,amssymb,bm} 
\usepackage{graphicx}  
\usepackage[tight]{subfigure}    
\usepackage{color} 
\usepackage[papersize={8.5in,11in}]{geometry}
\usepackage[colorlinks=true]{hyperref}  
\hypersetup{
    bookmarks=true,         
    unicode=false,          
    pdftoolbar=true,        
    pdfmenubar=true,        
    pdffitwindow=false,     
    pdfstartview={FitH},    
    pdftitle={Hyperscaling violation at the Ising-nematic quantum critical point in two dimensional metals},    
    pdfauthor={Andreas Eberlein, Ipsita Mandal, Subir Sachdev},     
    pdfsubject={Transport at quantum critical points},   
    pdfcreator={Creator},   
    pdfproducer={Producer}, 
    pdfkeywords={Ising-nematic quantum critical point} {Optical conductivity} {Free energy}, 
    pdfnewwindow=true,      
    colorlinks=true,       
    linkcolor=magenta, 
    citecolor=blue,        
    filecolor=magenta,      
    urlcolor=blue           
} 

\usepackage{amsfonts}
\usepackage{upgreek}
\usepackage{slashed}
\usepackage{latexsym}

\usepackage{todonotes}

\newcommand{\beq}{\begin{equation}}
\newcommand{\eeq}{\end{equation}}
\def\bea{\begin{eqnarray}}
\def\eea{\end{eqnarray}}
\newcommand{\nn}{\nonumber \\}

\newcommand{\bs}{\boldsymbol}

\newcommand{\tr}{\operatorname{tr}}

\newcommand{\intd}[2]{\int\negthinspace\negthickspace\frac{d^{#2}%
#1}{(2\pi)^{#2}}}
\newcommand{\intdef}[3]{\int_{#2}^{#3}\negthickspace\frac{d #1}{%
2\pi}}
\newcommand{\intdefnopi}[3]{\int_{#2}^{#3}\negthickspace d #1}

\begin{document}

\title{Hyperscaling violation at the Ising-nematic quantum critical point\\ in two dimensional metals}

\author{Andreas Eberlein}

\affiliation{Department of Physics, Harvard University, Cambridge MA 02138, USA}

\author{Ipsita Mandal}

\affiliation{Perimeter Institute for Theoretical Physics, Waterloo, Ontario, Canada N2L 2Y5}

\author{Subir Sachdev}

\affiliation{Department of Physics, Harvard University, Cambridge MA 02138, USA}
\affiliation{Perimeter Institute for Theoretical Physics, Waterloo, Ontario, Canada N2L 2Y5}

\date{\today\\
\vspace{0.6in}}

\begin{abstract}
Understanding optical conductivity data in the optimally doped cuprates in the framework of quantum criticality
requires  a strongly-coupled quantum critical metal which violates hyperscaling. In the simplest scaling framework, hyperscaling
violation can be characterized by a single non-zero exponent $\theta$, so that in a spatially isotropic state in $d$ spatial dimensions,
the specific heat scales with temperature as $T^{(d-\theta)/z}$, and the optical 
conductivity scales with frequency as $\omega^{(d-\theta-2)/z}$ for $\omega \gg 
T$, where $z$ is the dynamic critical exponent. We study the Ising-nematic 
critical point, using the controlled dimensional regularization method proposed 
by Dalidovich and Lee (Phys. Rev. B {\bf 88}, 245106 (2013)). We find that 
hyperscaling is violated, with $\theta =1$ in $d=2$. We expect that similar 
results apply to Fermi surfaces coupled to gauge fields in $d=2$.
\end{abstract}

\maketitle

\tableofcontents

\section{Introduction}
\label{sec:intro}

The widespread observation of `strange metal' behavior in numerous correlated electron compounds
underscores the need for a general theoretical framework for understanding metallic states without quasiparticle
excitations \cite{SSBK11}. Theories of such metallic states involve fermionic excitations across a Fermi surface coupled to low
energy and long-wavelength excitations of some gapless boson. This boson can either be a symmetry-breaking order 
parameter at a critical point \cite{Hertz76,Millis92,Chubukov00,OKF01,Metlitski2010a,MMSS10b,HHMS11,Dalidovich2013,Patel2015}, an emergent deconfined gauge 
field \cite{PAL89,lee94,Hermele04,SSL09,Metlitski2010a,MandalLee15}, and/or a critical `Higgs' field associated with phase transition
between different phases of a gauge theory \cite{Kaul08,DCSS15b}. In all of these cases, the critical theory of the non-quasiparticle metal can
be formulated as a continuum theory with an exactly conserved momentum density $\bs P$ \cite{HMPS,AAPSS14,DCSS15b}.
The other conserved quantities in such theories are the fermion number density and the energy density.

Such a continuum theory can provide a reliable computation for numerous single particle and other non-transport response functions. 
However, the conservation
of $\bs P$ leads to singularities in the transport properties which have to be regulated by various ``lattice'' contributions.
Umklapp scattering and/or impurities are needed to dissipate the momentum, and to obtain finite transport co-efficients in the
d.c. limit. At frequencies $\omega > T$, {\em e.g.\/} in the optical conductivity of interest in the present paper, 
the effects of $\bs P$ are less important; nevertheless, it is important to subtract out the singular contributions in the d.c. limit
to properly define the scaling properties of frequency-dependent transport co-efficients.
In a number of recent papers, `memory function', hydrodynamic, and holographic methods have been employed to understand 
the lattice contributions to the low frequency transport \cite{HKMS,Andreev11,HH12,LSS14,AL15,ALSS15,Lucas16}. 

For our purposes, it is useful to describe the transport properties in the limit where $\bs P$ is exactly conserved. Then the thermoelectric
response is described by
\beq
\left( \begin{array}{c}
{\bs J} \\ {\bs Q} \end{array} \right)  = \left( \begin{array}{cc} \sigma & \alpha \\
T \alpha & \overline{\kappa} \end{array} \right)
\left( \begin{array}{c}
{\bs E} \\  - {\bs \nabla} T \end{array} \right),
\label{intro1}
\eeq
where $T$ is temperature, ${\bs E}$ is an applied electric field, ${\bs J}$ is the electrical current, and ${\bf Q}$ is the heat current.
The electrical conductivity, $\sigma$, and thermoelectric conductivities $\alpha$, $\overline{\kappa}$, are in general spatial matrices,
but we will only consider here spatially isotropic systems without an external 
magnetic field, and then these conductivities are numbers.
The thermal conductivity, $\kappa$, is defined under conditions under which $\bs J = 0$, and so 
\beq
\kappa = \overline{\kappa} - \frac{T \alpha^2}{\sigma}.
\label{intro2}
\eeq

In systems with $\bs P$ conserved, the thermoelectric conductivities have poles at zero frequency, $\omega$, and obey \cite{ALSS15}
\bea 
\sigma &=& \frac{\mathcal{Q}^2}{\mathcal{M}}\, \left( \frac{1}{- i \omega} \right) + \sigma_Q   \nn
\alpha &=&  \frac{\mathcal{S} \mathcal{Q}}{\mathcal{M}} \, \left( \frac{1}{- i \omega} \right) + \alpha_Q  \nn
\overline{\kappa} &=&  \frac{T \mathcal{S}^2}{\mathcal{M}} \, \left( \frac{1}{- i \omega} \right) + \overline{\kappa}_Q,  
\label{intro3}
\eea
where $\sigma_Q$, $\alpha_Q$, $\overline{\kappa}_Q$ are the frequency-dependent conductivities after the pole has been subtracted out.
The residues of the pole are related exactly to static thermodynamic observables: these are the entropy density, $\mathcal{S}$, 
the current-momentum correlator $\mathcal{Q} \equiv \chi_{J_x,P_x}$, and the momentum-momentum correlator 
$\mathcal{M} \equiv \chi_{P_x, P_x}$. Combining Eqs.~(\ref{intro2}) and (\ref{intro3}), we observe that the pole at $\omega=0$
does not appear in $\kappa$, and in the d.c. limit \cite{HKMS,MBH13}
\beq
\kappa = \overline{\kappa}_Q - 2 \left( \frac{T \mathcal{S}}{\mathcal{Q}} \right) \alpha_Q
+ \left( \frac{T \mathcal{S}^2}{\mathcal{Q}^2} \right) \sigma_Q \quad , \quad \omega \rightarrow 0.
\label{intro4}
\eeq

In many cases, the $\sigma_Q$, $\alpha_Q$, $\overline{\kappa}_Q$ conductivities are not independent of each other, and obey identities connecting
them at all frequencies \cite{LHSRMP}. 
This is the case in systems with Hamiltonians which are invariant under relativistic or Galilean transformations.
However, our interest here is in systems which conserve $\bs P$, but do not enjoy relativistic or Galilean invariance, and such systems have not been
as extensively studied. In such situations, it appears that $\sigma_Q$, $\alpha_Q$, and $\overline{\kappa}_Q$ are independent response functions.

Let us now turn to the specific case of the Ising-nematic quantum critical point in two-dimensional metals \cite{OKF01,Metlitski2010a,Dalidovich2013}. Among the thermodynamic
observables introduced above, $\mathcal{Q}$ and $\mathcal{M}$ take constant non-critical values which depend upon  microscopic
details. However, the entropy density, $\mathcal{S}$ does have a singular $T$ dependence. From general scaling considerations, 
and allowing for violating of hyperscaling in which the spatial dimension $d \rightarrow d - \theta$, we expect \cite{HSS11}
\beq
\mathcal{S} \sim T^{(d-\theta)/z}, \label{intro5}
\eeq
with $z$ the dynamic critical exponent. We can view Eq.~(\ref{intro5}) as the definition of the value $\theta$.
In Section~\ref{sec:entropy},
we will use the controlled $\epsilon$-expansion for the Ising-nematic critical theory introduced by Dalidovich and Lee \cite{Dalidovich2013} 
to compute
$\mathcal{S}$. In $d=2$ we find the value
\beq
\theta = 1.
\label{intro6}
\eeq
Roughly speaking, this violation of hyperscaling can be traced to the fact that the momentum integral along the Fermi surface is non-singular,
and so only introduces an overall factor of the Fermi surface size. The single momentum dimension corresponding to this integral corresponds
to the value in Eq.~(\ref{intro6}).

For the frequency-dependent thermoelectric conductivities, similar scaling arguments \cite{Patel2015}, followed by $d \rightarrow d - \theta$ yield
\beq
\sigma_Q \sim \alpha_Q \sim \frac{\overline{\kappa}_Q}{T} \sim T^{(d-2-\theta)/z} \, \Upsilon (\omega /T),
\label{intro7}
\eeq
where $\Upsilon$ is a scaling function, and the three conductivities have separate scaling functions.
In Sections~\ref{sec:current}-\ref{sec:sigmaTg0}, we will use the Dalidovich-Lee $\epsilon$ expansion 
to compute $\sigma_Q$ in the regime $\omega \gg T$
(this is the `optical' conductivity). In this regime, and for $d=2$, we find $ \sigma_Q \sim \omega^{(d-2-\theta)/z}$, 
as expected from Eq.~(\ref{intro7}), with the value of $\theta$ again given by Eq.~(\ref{intro6}). 

We note that we have defined the value of $z$ by the scaling of the fermion response function transverse to the Fermi surface.
The Ising-nematic critical point has $z=3/2$ and $\theta=1$ in $d=2$, and so we have
$\sigma_Q \sim \omega^{-2/3}$. This scaling of the optical conductivity
was obtained earlier \cite{lee94} for the case of a Fermi surface coupled to a U(1) gauge field, but was given a different physical interpretation \cite{Patel2015}.


We will begin in Section~\ref{sec:action} by describing the action for the Ising nematic critical point.
The optical conductivity will be computed in Section~\ref{sec:current}, and the free energy and entropy density in Section~\ref{sec:entropy}.

\section{Action and scaling analysis at tree level}
\label{sec:action}

We consider a theory of fermions in $(2+1)$ dimensions which are coupled to a 
critical boson,
\begin{equation}
\begin{split}
	S(\bar\psi,\psi,\Phi) = &\sum_{s = \pm} \sum_{j = 1}^N \intd{k}{3} 
\tilde\psi^\dagger_{s j}(k) (i k_0 + s k_x + k_y^2) \tilde \psi_{sj}(k)\\
	& + \frac{1}{2} \intd{k}{3} (k_0^2 + k_x^2 + k_y^2) \Phi(-k) \Phi(k)\\
	& + \frac{e}{\sqrt{N}} \sum_{s = \pm} \sum_{j = 1}^N \intd{k}{3} \intd{q}{3} 
\lambda_s \Phi(q) \tilde\psi^\dagger_{s j}(k+q) \tilde \psi_{s j}(k),
\end{split}
\end{equation}
where $e$ is the fermion-boson coupling constant, $s = \pm 1$ labels the two 
Fermi surface patches and $\lambda_s$ equals 1 ($s$) for the Ising-nematic 
critical point (fermions coupled to a $U(1)$ gauge field). This model has been 
studied by many authors, including Refs.~\onlinecite{Metlitski2010a, 
Dalidovich2013}. In the following, we restrict ourselves to the Ising-nematic 
critical point and set $\lambda_s = 1$.

Introducing the spinor notation
\begin{align}
	\psi_j(k) &= \begin{pmatrix}
	             	\tilde \psi_{+,j}(k), \tilde \psi^\dagger_{-,j}(-k)
	             \end{pmatrix}^T		&%
	\bar\psi_j(k) &= \psi^\dagger_j(k) \gamma_0
\end{align}
with the gamma matrices $\gamma_0 = \sigma_y$ and $\gamma_x = \sigma_x$, the 
action can be rewritten as
\begin{equation}
	\begin{split}
		S(\bar\psi,\psi,\Phi) = & \sum_{j = 1}^N \intd{k}{3} \bar\psi_j(k) [i k_0 
\gamma_0 + i (k_x + k_y^2) \gamma_x] \psi_j(k)\\
	& + \frac{1}{2} \intd{q}{3} (q_0^2 + q_x^2 + q_y^2) \Phi(-q) \Phi(q)\\
	& + \frac{ie}{\sqrt{N}} \intd{k}{3}\intd{q}{3} \Phi(q) 
\bar\psi_j(k+q)\gamma_1 \psi_j(k).
	\end{split}
\end{equation}

In order to obtain a controlled perturbative expansion for correlation 
functions, we use the dimensional regularization proposed by Dalidovich and 
Lee~\cite{Dalidovich2013}, which increases the codimension of the Fermi 
surface. The dimensionally regularized action in $(d+1)$ dimensions reads
\begin{equation}
\begin{split}
	S(\bar\psi,\psi,\Phi) = & \sum_{j = 1}^N \intd{k}{d+1} \bar\psi_j(k) [i\bs 
\Gamma \cdot \bs K + i \gamma_x \delta_k] \psi_j(k)\\
& + \frac{1}{2} \intd{q}{d+1} [\bs Q^2 + q_x^2 + q_y^2] \Phi(-q) \Phi(q)\\
& + \frac{i e}{\sqrt{N}} \sqrt{d-1} \sum_{j = 1}^N \intd{k}{d+1} \intd{q}{d+1} 
\Phi(q) 
\bar\psi_{j}(k+q) \gamma_x \psi_j(k),
\end{split}
\end{equation}
where $\bs K = (k_0, k_1, \ldots, k_{d-2})$ represents frequency and $(d-2)$ 
components of the full $(d+1)$-dimensional energy-momentum vector. $k_1$, 
\ldots, $k_{d-2}$ are the time-like auxiliary dimensions. The 
gamma matrices for the new dimensions are $\bs \Gamma = (\gamma_0, \gamma_1, 
\ldots, \gamma_{d-2})$. We introduced the abbreviation $\delta_k = k_x + 
\sqrt{d-1} k_y^2$ and keep the definitions $\gamma_0 = \sigma_y$ and 
$\gamma_x = \sigma_x$.

Rescaling momenta as
\begin{align}
	\bs K &= b^{-1} \bs K'		&		k_x &= b^{-1} k_x'		& k_y = b^{-1/2} 
k_y',
\end{align}
the fermionic quadratic part of the action is invariant under rescaling for
\begin{align}
	\psi_j(k) = b^{d/2 + 3/4} \psi_j'(k').
\end{align}
Rescaling the bosonic fields as
\begin{align}
	\Phi(k) = b^{d/2 + 3/4} \Phi'(k')
\end{align}
the term $\sim q_y^2$ in the bosonic quadratic part is invariant under 
rescaling while the terms proportional to $\bs Q^2$ and $q_x^2$ are 
irrelevant. The interaction part changes under rescaling like
\begin{equation}
	e' = e b^{\frac{1}{2}(5/2-d)},
\end{equation}
identifying $d = 5/2$ as the upper critical dimension. The coupling $e$ is 
irrelevant for $d > 5/2$ and relevant for $d < 5/2$. This allows to access 
non-Fermi liquid physics perturbatively by using $\epsilon = 5/2 - d$ as 
expansion parameter.

Keeping only marginal terms, the ansatz for the local field theory reads
\begin{equation}
\begin{split}
	S(\bar\psi,\psi,\Phi) = &\sum_{j = 1}^N \intd{k}{d+1} \bar\psi_j(k) [i 
\bs\Gamma \cdot \bs K + i \gamma_x\delta_k] \psi_j(k) + \frac{1}{2} 
\intd{k}{d+1} k_y^2 \Phi(-k) \Phi(k)\\
& + \frac{i e \mu^{\epsilon/2}}{\sqrt{N}} \sqrt{d-1} \sum_{j = 1}^N 
\intd{k}{d+1} 
\intd{q}{d+1} \Phi(q) \bar\psi_j(k+q) \psi_j(k),
\end{split}
\label{eq:action}
\end{equation}
where we introduced the momentum scale $\mu$ in order to make the coupling $e$
dimensionless. Perturbative corrections to this action at one-loop level 
reintroduce dynamics for the bosonic field. The bare propagators read
\begin{gather}
	D_0(q) = \frac{1}{q_y^2}\\
	G_0(k) = \frac{\bs \Gamma \cdot \bs K + \gamma_x \delta_k}{i (\bs K^2 + 
\delta_k^2)}.
\end{gather}

\section{Current-current correlation function and optical conductivity}
\label{sec:current}
In this section, we compute the optical conductivity $\sigma(\omega) = 
\sigma_{xx}(\omega, \bs q = \bs 0)$ at $T = 0$ via the Kubo formula,
\begin{equation}
	\sigma(\omega) = -\frac{1}{\Omega_m} \langle J_x 
J_x\rangle(i\Omega_m)|_{i\Omega_m \rightarrow \omega + i0^+}.
\label{eq:Kubo}
\end{equation}
The current operator $J_x$ is obtained by minimally coupling the 
action in Eq.~\eqref{eq:action} to a vector potential that is non-zero only in 
the $x$-direction. We obtain
\begin{equation}
	J_x(x) = \sum_{j = 1}^N \bar\psi_j(x) i e_A \gamma_x \psi_j(x)
\label{eq:Gamma0}
\end{equation}
where we termed the charge $e_A$ in order to distinguish it from the coupling 
constant in the action.

Note that the current operator in Eq.~\eqref{eq:Gamma0} describes a ``chiral'' 
current. The current operator for the particle current is given by
\begin{equation}
	J_x^N(x) = \sum_{j = 1}^N \bar\psi_j(x) e_N \gamma_0 \psi_j(x).
\end{equation}
The chiral current is more convenient to use within codimensional 
regularization. In the physical dimension $d = 2$, the correlation functions 
for both currents are the same within the fixed point theory of Dalidovich and 
Lee~\cite{Dalidovich2013}. They differ at the three-loop 
level~\cite{Metlitski2010a} where both patches are coupled.

In the following sections we compute the current-current correlation function 
in $\mathcal O(\epsilon)$ and subsequently determine its scaling behaviour.

\begin{figure}
	\centering
	\subfigure[]{\includegraphics[width=0.3\textwidth]{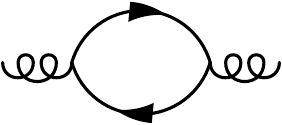} \label{fig:1a}}
	\subfigure[]{\includegraphics[width=0.3\textwidth]{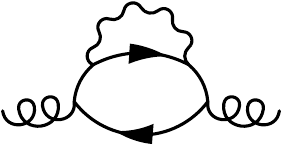} \label{fig:1b}}
	\subfigure[]{\includegraphics[width=0.3\textwidth]{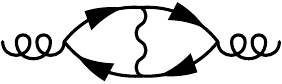} \label{fig:1c}}
	\caption{Feynman diagrams for the contributions to the current-current 
correlation function in (a) $\mathcal O(N)$ and (b,c) $\mathcal O(1)$. The 
wiggly line represents the bosonic propagator and the curly line the vector 
potential.}
	\label{fig1}
\end{figure}

\subsection{Current-current correlation function at one-loop level}
The current-current correlation function at 
one-loop level for $q = \omega \bs e_0 = \bs Q$ is given by a simple fermionic 
loop with two current insertions, as shown in Fig.~\ref{fig:1a},
\begin{equation}
\begin{split}
	\langle J_x J_x \rangle_\text{1loop}(i\omega) &= e_A^2 
\sum_{j = 1}^N \intd{k}{d+1} \tr\bigl(\gamma_x G_0(k+q) \gamma_x 
G_0(k)\bigr)\\
&= -2 e_A^2 N \intd{k}{d+1} \frac{\delta_k^2 - \bs K \cdot (\bs K + \bs Q)}{(\bs 
K^2 + \delta_k^2) ((\bs K + \bs Q)^2 + \delta_k^2)}.
\label{eq:jj_1Loop}
\end{split}
\end{equation}
Evaluation as described in Appendix~\ref{app:jj_free} yields for $d = 
\tfrac{5}{2} - \epsilon$
\begin{equation}
	\langle J_x J_x \rangle_\text{1loop} (i\omega) = -e_A^2 N 
\intdef{k_y}{}{} u_{\text{1Loop},\epsilon=0} |\omega|^{1/2-\epsilon},
\label{eq:JJ_1Loop}
\end{equation}
where
\begin{equation}
	u_{\text{1Loop},\epsilon=0} = \frac{\Gamma(\frac{5}{4})}{3\pi^{3/4}} \approx 
0.128038.
\label{eq:u1Loop}
\end{equation}
For $\epsilon = 1/2$, the one-loop result is independent of frequency, as 
expected.

\subsection{Two-loop self-energy correction to current-current correlation 
function}
The self-energy correction to the current-current correlation function at 
two-loop level for $q = \omega \bs e_0 = \bs Q$ reads
\begin{align}
	\langle J_x& J_x \rangle_\text{SE}(i\omega) = 2 e_A^2 \sum_{j=1}^N 
\intd{k}{d+1}\operatorname{tr}\Bigl(\gamma_xG_0(k+q)\gamma_x 
G_0(k)\Sigma_1(k) G_0(k)\Bigr)\\
	&= 4 e^{4/3} e_A^2 
\alpha_{\Sigma,d} \intd{k}{d+1} \Bigl(\frac{\mu}{|\bs 
K|}\Bigr)^\frac{2\epsilon}{3} \frac{2 \delta_k^2 \bs K^2 + \bs K 
\cdot (\bs K + \bs Q)(\delta_k^2 - \bs K^2)}{\bigl((\bs K + \bs Q)^2 + 
\delta_{k+q}^2\bigr)(\bs K^2 + \delta_k^2)^2},
\label{eq:jj_selfenergy1}
\end{align}
where $\Sigma_1(k)$ is the fermionic self-energy at one-loop level, 
Eq.~\eqref{eq:FermionicSelfenergy}. The self-energy correction is shown 
diagrammatically in Fig.~\ref{fig:1b}. This contribution contains a pole in 
$\epsilon^{-1}$ and evaluation as described in Appendix~\ref{app:jj_selfenergy} 
yields
\begin{equation}
	\langle J_x J_x \rangle_\text{SE}(i\omega) = e_A^2 e^{4/3} 
\epsilon^{-1} \intd{k_y}{} |\omega|^{\frac{1}{2} - \epsilon} 
\Bigl(\frac{\mu}{|\omega|}\Bigr)^{2\epsilon/3} a_{\Sigma,\epsilon=0} + \mathcal 
O(\epsilon^0),
\end{equation}
where we set $\epsilon = 0$ in the numerical prefactor,
\begin{equation}
a_{\Sigma,\epsilon=0} = \frac{\pi^{1/4} 
u_{\Sigma,\epsilon=0}}{8\sqrt{2} \Gamma(\frac{7}{4})} \approx 0.0086875.
\label{eq:aSigma2Loop}
\end{equation}

\subsection{Two-loop vertex correction to current-current correlation 
function}
\label{sec:2LVC}
The two-loop vertex correction contribution to the current-current correlation 
function, which is shown diagrammatically in Fig.~\ref{fig:1c}, is given by
\begin{align}
	\langle J_x& J_x\rangle_\text{VC}(i\omega) = -i e_A \sum_{j = 1}^N 
\intd{k}{d+1} \operatorname{tr}\bigl(\gamma_x G_0(k+q) \Gamma_1(\bs K, i 
\omega) G_0(k)\bigr)\\
	&= -e_A^2 N \intd{k}{d+1} \frac{\operatorname{tr}\bigr[(\bs \Gamma \cdot \bs 
K + \gamma_x 
\delta_k)\gamma_x(\bs \Gamma\cdot (\bs K + \bs Q) + \gamma_x 
\delta_k)\gamma_x\tilde\Gamma_1(\bs K, i\omega)\bigr]}{(\bs K^2 + 
\delta_k^2)((\bs K + \bs Q)^2 + \delta_k^2)}.
\label{eq:jj_2Loop_VC}
\end{align}
for $q = \omega \bs e_0 = \bs Q$. The vertex correction to the current vertex 
at one-loop level, $\Gamma_1(\bs K, \omega) = i e_A \gamma_x \tilde \Gamma_1(\bs 
K, \omega)$, is derived in Appendix~\ref{sec:jj_vertex1Loop} and 
given by Eq.~\eqref{eq:OneLoopCurrentVertex}. Evaluation of 
Eq.~\eqref{eq:jj_2Loop_VC} as described in Appendix~\ref{app:jj_vertex} yields a 
result that is free of poles in $\epsilon^{-1}$. Setting $\epsilon = 0$ in the 
numerical prefactors, we obtain
\begin{equation}
\langle J_x J_x\rangle_\text{VC} (i\omega) = - 
\alpha_\text{VC}^{\epsilon = 0} e_A^2 e^{4/3} 
|\omega|^{\frac{1}{2} - \epsilon} 
\Bigl(\frac{\mu}{|\omega|}\Bigr)^{2\epsilon/3} \intdef{k_y}{}{}
\end{equation}
where $\alpha_\text{VC}^{\epsilon = 0} \approx 0.0230903$.

\subsection{Scaling behavior of optical conductivity and free energy}
\label{sec:scaling}
In this section we determine the scaling behaviour of the optical conductivity, 
first from general scaling arguments and subsequently for the fixed point 
theory for the Ising-nematic quantum-critical point using the above results for 
the current-current correlation function.

\subsubsection{Scaling behavior of optical conductivity and free energy: 
General arguments}
\label{subsec:SalingConductivityGeneral}
In a system with spatial dimension $d$, dynamical critical exponent $z$, 
$1/2-\epsilon$ time-like auxiliary dimensions and violation of hyperscaling 
exponent $\theta$, the free energy has scaling dimension
\begin{equation}
	[F] = d-\theta + z + (1/2 - \epsilon)z = d-\theta + (3/2 - \epsilon) z.
\end{equation}
The current operator is given by $J = \frac{\delta F}{\delta A}$, 
where $A$ is the vector potential with scaling dimension one, and scales as $[J] 
= d-\theta-1 + (3/2 - \epsilon)z$. From the Kubo formula Eq.~\eqref{eq:Kubo}, 
we obtain the scaling dimension of the optical conductivity,
\begin{equation}
[\sigma] = -z - (d-\theta) - (3/2 - \epsilon) z + 2 [J] = d - \theta - 2 + (1/2 
- \epsilon) z,
\end{equation}
where we took into account that the number of spatial dimensions is effectively 
reduced by $\theta$. 

In $d = 2$, the free energy and optical conductivity thus scale as
\begin{align}
	F(T) &\sim T^{(2-\theta)/z + 3/2-\epsilon},		&		\sigma(\omega) &\sim 
\omega^{-\theta/z + 1/2 - \epsilon}.
\end{align}
In the $\epsilon$ expansion, it is expected \cite{Dalidovich2013} that
\beq
z = \frac{3}{3 - 2 \epsilon}.
\eeq	
In a system with the hyperscaling property and $\theta = 0$, we expect
\begin{gather}
	F(T) \sim T^{7/2 - 7\epsilon/3}		\\
	\sigma(\omega) \sim \omega^{1/2-\epsilon}.
\end{gather}
If hyperscaling is violated, the free energy and optical conductivity are 
expected to scale as
\begin{gather}
	F(T) \sim T^{5/2 - 5\epsilon/3}	\label{Fepsilon}	\\
	\sigma(\omega) \sim \omega^{-1/2-\epsilon/3}
\label{sigmaepsilon}
\end{gather}
for $\theta = 1$. Note that there are not expected to be any corrections to 
Eq.~\eqref{Fepsilon} and~\eqref{sigmaepsilon} at higher orders in $\epsilon$.
In a perturbative expansion in $\epsilon$, the result for the free 
energy and optical conductivity would behave like
\begin{gather}
	F(T) \sim T^{5/2 - \epsilon} \bigl(1 - (1 - 1/z) \operatorname{ln} T + 
\ldots\bigr) \label{eq:F_epsilon_01}\\
	\sigma(\omega) \sim \omega^{-1/2-\epsilon} \bigl(1 + (1 - 1/z) 
\operatorname{ln}\omega + \ldots\bigr),
\label{eq:sigma_epsilon_O1}
\end{gather}
where $1 - 1/z = 2\epsilon / 3$ for the above-mentioned fixed point theory.

\subsubsection{Scaling behavior of conductivity: Evaluation for fixed point 
theory}
The two-loop vertex correction contribution computed in Sec.~\ref{sec:2LVC} 
turned out to be finite, so that only the self-energy correction yields a 
renormalization of the scaling-behavior of the conductivity. The current-current 
correlation function is thus given by
\begin{gather}
	\langle J_x J_x\rangle (i\omega) \approx \langle J_x J_x\rangle_\text{1Loop} 
(i\omega) + \langle J_x J_x\rangle_\text{SE} (i\omega) + \ldots \\
	\langle J_x J_x\rangle_\text{1Loop} (i\omega) = -e_A^2 N 
\intdef{k_y}{}{} u_{\text{1Loop},\epsilon=0} |\omega|^{1/2-\epsilon}\\
	\langle J_x J_x\rangle_\text{SE} (i\omega) = e_A^2 e^{4/3} 
\epsilon^{-1} \intdef{k_y}{}{} |\omega|^{1/2-\epsilon} 
\Bigl(\frac{\mu}{|\omega|}\Bigr)^{2\epsilon/3} a_{\Sigma,\epsilon=0}+\ldots,
\end{gather}
where the last line contains only the pole contribution. Resummation yields
\begin{align}
	\langle J_x J_x\rangle (i\omega) &= -e_A^2 N 
\intdef{k_y}{}{} u_{\text{1Loop},\epsilon=0} |\omega|^{1/2-\epsilon} \Bigl\{1 - 
\frac{e^{4/3}}{N\epsilon} \Bigl(\frac{\mu}{|\omega|}\Bigr)^{2\epsilon/3} 
\frac{a_{\Sigma,\epsilon=0}}{u_{\text{1Loop},\epsilon=0}}\Bigr\}\\
& \approx -e_A^2 N \intdef{k_y}{}{} u_{\text{1Loop},\epsilon=0} 
|\omega|^{1/2-\epsilon} \Bigl\{1 + \gamma
\operatorname{ln}\Bigl(\frac{|\omega|}{\mu}\Bigr)\Bigr\} 
\label{eq:jj_epsilon_O1}\end{align}
where
\begin{equation}
	\gamma = \frac{2}{3} 
\frac{e^{4/3}}{N} \frac{a_{\Sigma,\epsilon=0}}
{u_{\text{1Loop},\epsilon=0}}.
	\label{eq:exponent}
\end{equation}
The coupling $\frac{e^{4/3}}{N}$ is evaluated at the fixed 
point~\cite{Dalidovich2013} using the $\beta$-function in $\mathcal 
O(\epsilon)$, yielding
\begin{equation}
	\Bigl(\frac{e^{4/3}}{N}\Bigr)^\ast = u^{-1}_{\Sigma,\epsilon=0} \epsilon.
\end{equation}
Inserting this result together with 
$u_{\text{1Loop},\epsilon=0}$ and $a_{\Sigma,\epsilon=0}$ from 
Eq.~\eqref{eq:u1Loop} and Eq.~\eqref{eq:aSigma2Loop}, respectively, into 
Eq.~\eqref{eq:exponent}, we indeed obtain the value
\beq
\gamma = \frac{2\epsilon}{3},
\eeq
which is expected from Eq.~(\ref{eq:sigma_epsilon_O1}) for $\theta = 1$.

\subsection{Pole contribution to conductivity}
\label{sec:current-momentum}
In Sec.~\ref{sec:intro}, we argued that the conductivity typically consists of 
a pole contribution and a ``quantum'' contribution $\sigma_Q$. In the results 
of the last sections, no pole contribution appeared. In order to understand 
this better, we compute the current-momentum susceptibility in the following. 
It is given by
\begin{equation}
	\chi_{J_x, P_x} = \lim_{\bs q \rightarrow 0} \langle J_x 
P_x\rangle (q_0 = 0, \bs q).
\end{equation}
At one-loop level and for $\bs q \neq \bs 0$, it reads
\begin{equation}
\begin{split}
	\langle J_x P_x&\rangle_\text{1Loop} (q_0 = 0, \bs q) = -i e_A N 
\intd{k}{d+1} \bigl(k_x + \frac{q_x}{2}\bigr) \operatorname{tr}\bigl(\gamma_x 
G_0(k+q) \gamma_0 G_0(k)\bigr)\\
	&= i e_A N 
\intd{k}{d+1} \bigl(k_x + \frac{q_x}{2}\bigr) \frac{\operatorname{tr}\bigl\{ 
\gamma_x [\bs \Gamma\cdot \bs K + \gamma_x \delta_{k+q}]\gamma_0 [\bs 
\Gamma\cdot \bs K + \gamma_x \delta_k]\bigr\}}{(\bs K^2 + \delta_{k+q}^2) (\bs 
K^2 + \delta_k^2)}
\label{eq:ChirCurrentMomSusceptibility}
\end{split}
\end{equation}
where
\begin{equation}
	P_x(x) = \frac{i}{2} \sum_{j = 1}^{N} \Bigl(\bar\psi_j(x) \gamma_0 
\partial_x \psi_j(x) - \partial_x \bar\psi_j(x) \gamma_0 \psi_j(x)\Bigr)
\end{equation}
is the $x$-component of the momentum density operator associated with the 
physical time direction. Computing the trace over gamma matrices,
\begin{equation}
\operatorname{tr}\bigl\{ \gamma_x [\bs 
\Gamma\cdot \bs K + \gamma_x \delta_{k+q}]\gamma_0 [\bs \Gamma\cdot \bs K + 
\gamma_x \delta_k]\bigr\} = 2 K_0 (\delta_k + \delta_{k+q}),
\end{equation}
and inserting the result into Eq.~\eqref{eq:ChirCurrentMomSusceptibility}, it 
is easy to see that $\chi_{J_x, P_x}$ vanishes at one-loop level because the 
integrand is an odd function of $K_0$. 

This result is also expected to hold beyond one-loop level, because the 
charge associated with the ``chiral'' current $J_x$ measures the difference 
between the occupation numbers at the two opposite patches of the Fermi surface 
and vanishes. Moreover, in $d = 2$, $J_x$ is equal to the fermionic density 
operator of the model. In that case, $\chi_{J_x, P_x} = \lim_{\bs q \rightarrow 
\bs 0} \langle J_x P_x \rangle (q_0 = 0, \bs q)$ is a correlation function 
between an operator that is odd under time reversal or spatial inversion 
($P_x$) and one that is even under these symmetries ($J_x$) and thus has to 
vanish. The conductivity computed in the last section is thus the ``quantum'' 
contribution $\sigma_Q$. 

The optical conductivity for the particle current in $d = 2$ consists 
of the same quantum contribution $\sigma_Q$ and an additional pole 
contribution. The presence of the pole contribution follows from the fact that 
the particle current-momentum susceptibility,
\begin{equation}
	\chi_{J^N P} = \lim_{\bs q \rightarrow 0} \langle J^N_x 
P_x\rangle (q_0 = 0, \bs q),
\end{equation}
is non-zero in $d = 2$. This is shown in Appendix~\ref{sec:JCPsusceptibility}, 
where we obtain
\begin{gather}
\langle J_x^N P_x\rangle_\text{1Loop}(0) = -\frac{e_N N}{\pi} \intd{k_y}{} 
k_y^2.
\end{gather}
at one-loop level.

\section{Free energy at finite temperature}
\label{sec:entropy}
In this section we compute the free energy at finite temperature in order to 
study the $T > 0$ dynamics at the Ising-nematic QCP. For this purpose we have 
to compute the contributions of free fermions, free bosons and a self-energy 
correction due to their interaction. 

\subsection{Contribution of free fermions}
The contribution of free fermions is given by
\begin{equation}
	F_{f,0}(T) - F_{f,0}(0) = -\intd{k}{2} \intd{K'}{1/2-\epsilon} 
\Bigl[T\sum_{n = \pm}\ln\bigl(1 + e^{- n \sqrt{\bs K'^2 + \delta_k^2} / 
T}\bigr) - \sqrt{\bs K'^2 + \delta_k^2}\Bigr],
\end{equation}
where we subtracted the result at $T = 0$ in order to make it finite. 
Shifting $k_x \rightarrow k_x - \sqrt{d-1} k_y^2$ and rescaling $\bs K' 
\rightarrow T \bs K'$, $k_x \rightarrow T k_x$ we obtain
\begin{gather}
	=-T^d \intd{k}{2} \intd{K'}{1/2-\epsilon}\Bigl[\ln\bigl(1 + 
e^{\sqrt{\bs K'^2 + k_x^2}}\bigr) + \ln\bigl(1 + 
e^{-\sqrt{\bs K'^2 + k_x^2}}\bigr) - \sqrt{\bs K'^2 + k_x^2}~\Bigr]\\
 = - T^{5/2-\epsilon} 
\intd{k_y}{} \alpha_{f,0},
\end{gather}
where
\begin{equation}
\begin{split}
	\alpha_{f,0}& = \frac{S_{3/2}}{(2\pi)^{3/2}} \intdefnopi{p}{0}{\infty}\, 
\sqrt{p} \, \Bigl[\ln\bigl(1 + e^p\bigr) + \ln\bigl(1 + e^{-p}\bigr) - p\Bigr] = 
\frac{(2\sqrt{2} - 1) \Gamma(\frac{5}{4}) \zeta(\frac{5}{2})}{\sqrt{2} 
\pi^{5/4}} \approx 0.375866
\end{split}
\end{equation}
after setting $\epsilon = 0$ in the numerical prefactor.

\subsection{Contribution of free bosons}
We compute the contribution of free bosons to the free energy for an 
inverse bare propagator that is quadratic in all frequency and momentum 
arguments and obtain
\begin{align}
	F_{b,0}(T) = T \intd{q}{2} \intd{Q'}{\frac{1}{2}-\epsilon} \ln\Bigl(1 - 
e^{-\sqrt{\bs Q'^2 + 
\bs q^2}/T}\Bigr) = -T^{7/2 - \epsilon} \alpha_{b,0}.
\end{align}
In the last step we set $\epsilon = 0$ in the numerical prefactor and defined
\begin{equation}
	\alpha_{b,0} = -\frac{S_{5/2}}{(2\pi)^{5/2}} \intdefnopi{p}{0}{\infty}\, 
p^{3/2} \ln\bigl(1 - e^{-p}\bigr) = \frac{3\zeta(\frac{7}{2})}{8\sqrt{2} 
\pi^{3/4} \Gamma(\frac{5}{4})} \approx 0.139686.
\end{equation}

\subsection{Interaction correction to the fermionic part of the free energy}
\label{subsec:FreeEnergyIntCorrectionFermions}
\begin{figure}
	\centering
	\includegraphics[width=0.15\linewidth]{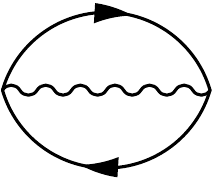}
	\caption{Feynman diagram for the two-loop interaction correction to the free 
energy.}
	\label{fig2}
\end{figure}

The lowest-order interaction correction to the free energy is given by
\begin{equation}
\begin{split}
	F_{f,b}(T) &= -\frac{e^2\mu^\epsilon (d-1)}{2N} 
T\sum_{\Omega_m}\intd{q}{2} \intd{Q'}{\frac{1}{2}-\epsilon} 
T\sum_{\omega_n}\intd{k}{2} 
\intd{K'}{\frac{1}{2}-\epsilon} D_1(\Omega_m, \bs Q', \bs q) \\
	&\qquad \times \operatorname{tr}\Bigl[\gamma_xG_0(\omega_n + \Omega_m, 
\bs K' + \bs Q', \bs k + \bs q) \gamma_x G_0(\omega_n, \bs K', \bs 
k)\Bigr],
\end{split}
\end{equation}
where $\omega_n$ and $\Omega_m$ are fermionic and bosonic Matsubara 
frequencies, respectively. The corresponding Feynman diagram is shown in 
Fig.~\ref{fig2}. From this expression, we need to isolate the pole 
contributions. In lowest order in $\epsilon$, these are obtained by evaluating 
one frequency sum as an integral in the limit $T \rightarrow 0$ and the other 
one at finite temperature. In case the continuous frequency appears in the 
argument of the bosonic and a fermionic propagator, we can rewrite the diagram 
as a fermionic loop with an 
insertion of the fermionic self-energy at $T = 0$. Note that there are two such 
contributions. In case the continuous frequency variable appears in the two 
fermionic propgators, we can replace them by the bosonic self-energy at $T = 0$.

The interaction correction to the fermionic part of the free energy is then 
given by
\begin{equation}
\begin{split}
	F_{f,b}^{(1)}(T) &= -\frac{e^2\mu^\epsilon (d-1)}{N} \intd{Q_0}{} 
\intd{q}{2} \intd{Q'}{\frac{1}{2}-\epsilon} \intd{k}{2} 
\intd{K'}{\frac{1}{2}-\epsilon} T\sum_{\omega_n} D_1(Q_0, \bs Q', \bs q) \\
	&\qquad \times \operatorname{tr}\Bigl[\gamma_xG_0(\omega_n + Q_0, \bs K' 
+ \bs Q', \bs k + \bs q) \gamma_x G_0(\omega_n, \bs K', \bs k)\Bigr]\\
	&= \intd{k}{2}\intd{K'}{\frac{1}{2}-\epsilon} T \sum_{\omega_n} 
\operatorname{tr}\bigl[\Sigma_1(\omega_n, \bs K', \bs k) G_0(\omega_n, \bs K', 
\bs k)\bigr],
\end{split}
\end{equation}
where $\Sigma_1$ is the fermionic self-energy at $T = 0$, 
Eq.~\eqref{eq:FermionicSelfenergy}, and we included a 
factor of two for the two possibilities of obtaining the self-energy insertion.

After the computation of the trace over gamma matrices and a shift of $k_x 
\rightarrow k_x - \sqrt{d-1} k_y^2$, this contribution reads
\begin{equation}
\begin{split}
	F_{f,b}^{(1)}(T) &= -\frac{2e^{4/3}}{N} \alpha_{\Sigma,d} 
\mu^{2\epsilon/3} \intd{k}{2} \intd{K'}{\frac{1}{2}-\epsilon} T \sum_{\omega_n} 
\frac{1}{(\omega_n^2 + \bs K'^2)^{2\epsilon / 6 - 1} (\omega_n^2 + \bs K'^2 + 
k_x^2)}\\
	&= -\alpha_{f,1} \frac{e^{4/3}}{N} \intd{k_y}{} \epsilon^{-1} 
			\Bigl(\frac{\mu}{T}\Bigr)^{2\epsilon/3} T^{5/2-\epsilon},
\end{split}
\end{equation}
where in the last step we computed the integrals using Feynman 
parameters and defined $\alpha_{f,1} = \alpha_{f,0} u_{\Sigma,0}$. Combining 
this result with the free fermion contribution yields
\begin{equation}
\begin{split}
	F_{f,0}(T) - F_{f,0}(0) + F_{f,b}^{(1)}(T) &= -\alpha_{f,0} \intd{k_y}{} 
T^{5/2-\epsilon} - \alpha_{f,1} \frac{e^{4/3}}{N} \intd{k_y}{} \epsilon^{-1} 
\Bigl(\frac{\mu}{T}\Bigr)^{2\epsilon/3} T^{5/2-\epsilon} \\
	&= -\alpha_{f,0} \intd{k_y}{} T^{5/2-\epsilon} \Bigl(1 - \gamma_f \ln 
\frac{T}{\mu}\Bigr),
\end{split}
\end{equation}
where
\begin{equation}
	\gamma_f = \frac{2}{3} \frac{\alpha_{f,1}}{\alpha_{f,0}} \frac{e^{4/3}}{N}.
\end{equation}
Evaluating $\gamma_f$ at the fixed point and exploiting 
$(\frac{e^{4/3}}{N})^\ast = \frac{\epsilon}{u_{\Sigma,0}}$, we obtain 
$\gamma_f^\ast = \frac{2\epsilon}{3}$ and thus
\begin{equation}
	F_{f,0}(T) - F_{f,0}(0) + F_{f,b}^{(1)}(T) = -\alpha_{f,0} \intd{k_y}{} 
T^{5/2-\epsilon} \Bigl(1 - \frac{2\epsilon}{3} 
\ln\frac{T}{\mu}\Bigr).
\end{equation}
This temperature dependence is expected in this order in $\epsilon$ from 
Eq.~\eqref{eq:F_epsilon_01} in case hyperscaling is violated with $\theta = 1$ 
for $d = 2$.

\subsection{Interaction correction to the bosonic part of the free energy}
\label{subsec:FreeEnergyIntCorrectionBosons}
In this section we evaluate the interaction correction to the bosonic part of 
the free energy. In order to extract the leading order contribution in 
$\epsilon$, the bosonic self-energy entering the diagram as self-energy 
insertion and in the bosonic propagator is evaluated at $T = 0$, while the 
remaining Matsubara frequency sum is evaluated at $T > 0$. This yields
\begin{equation}
\begin{split}
	F_{f,b}^{(2)}(T) &= -\frac{T}{2} \intd{q}{2} \intd{Q'}{\frac{1}{2}-\epsilon} 
\sum_{\Omega_m} 
D_1(\Omega_m, \bs Q', \bs q) \Pi(\Omega_m, \bs Q', \bs q) \\
	& = \frac{T}{2} \beta_d e^2 \mu^\epsilon (d-1) \intd{q}{2} 
\intd{Q'}{\frac{1}{2}-\epsilon} 
\sum_{\Omega_m} \frac{\sqrt{\Omega_m^2 + \bs Q'^2}^{d-1}}{|q_y|^3 + \beta_d e^2 
\mu^\epsilon \sqrt{\Omega_m^2 + \bs Q'^2}^{d-1}}.
\end{split}
\end{equation}
Computing the integrals and evaluating the frequency sum using zeta-function 
regularization identities~\cite{Patel2015}, we obtain
\begin{align}
	&= \frac{e^{2/3} (\beta_d \mu^\epsilon)^{1/3} S_{d-2} (d-1)}{3 \sqrt{3} 
(2\pi)^{d-2}} T \sum_m \intd{q_x}{} \intdefnopi{Q}{0}{\infty} Q^{d-3} 
\sqrt{\Omega_m^2 + Q^2}^\frac{d-1}{3}\\
	&= \frac{\Gamma(\frac{1}{4} - \frac{\epsilon}{2})\Gamma(-\frac{1}{2} + 
\frac{2\epsilon}{3}) S_{d-2} (d-1)}{6 \sqrt{3} \Gamma(\frac{\epsilon}{6} - 
\frac{1}{4}) (2\pi)^{d-2}} e^{2/3} (\beta_d \mu^\epsilon)^{1/3} 
\intd{q_x}{} T \sum_m \frac{1}{|\Omega_m|^{4\epsilon/3 - 1}}\\
	&= -\frac{\pi^{5/4} (\beta_{5/2})^{1/3}}{12 
\sqrt{6} \Gamma(\frac{3}{4})} \intd{q_x}{} e^{2/3} T^{2-\epsilon} 
\Bigl(\frac{\mu}{T}\Bigr)^{\epsilon/3},
\end{align}
where we set $\epsilon = 0$ in the numerical prefactor.

In this order of approximation, the interaction correction to the bosonic part 
of the free energy does not contain a pole in $\epsilon$ and the bosonic 
contribution to the free energy is thus not renormalized.

\section{Conclusions}

We have computed the optical conductivity and the free energy at the 
Ising-nematic quantum critical point in two-dimensional metals using the 
$\epsilon$-expansion introduced by Dalidovich and Lee~\cite{Dalidovich2013}. 
This method allows to study the non-Fermi liquid regime at this strongly 
coupled critical point in a controlled way as a stable fixed point of the 
renormalization group flow. We found that hyperscaling is 
violated with a violation of hyperscaling exponent $\theta = 1$ in $d = 2$.

The optical conductivity scales as $\sigma(\omega) \sim \omega^{-2/3}$ at the 
fixed point, which is close to the behaviour found in optimally doped 
cuprates~\cite{Marel2003}. This scaling behaviour of the optical conductivity 
was obtained before in Ref.~\cite{lee94} for a metal coupled to a $U(1)$ gauge 
field, but was given a different physical interpretation~\cite{Patel2015}.

We also computed the free energy at finite temperature, $T > 0$. The results 
for the fermionic contribution to the free energy confirm violation of 
hyperscaling with the same exponent $\theta = 1$ in $d = 2$. At lowest order in 
$\epsilon$, the bosonic contribution to the free energy was not renormalized.

In critical points without disorder, the violation of hyperscaling has previously been associated
with systems above their upper-critical dimension, where the critical theory is essentially
a free field theory \cite{MEF73}. As far as we are aware, our computation in the present paper
is the first to systematically demonstrate violation of hyperscaling at a strongly-coupled fixed point. 
The origin of the violation was the presence of a Fermi surface, and the independence of the singular
terms on the momentum direction parallel to the Fermi surface. A previous computation in a system
with a Fermi surface \cite{Patel2015}, which was dominated by singular contributions 
at hot spots on the Fermi surface, instead found that hyperscaling was preserved.
We believe that with hyperscaling violation established, the path is open in similar models
to understand the anomalous optical conductivity of strange metals \cite{Marel2003}.

\section*{Acknowledgments} 

We would like to thank A.~A.~Patel for valuable discussions. 
This research was supported by the NSF under Grant DMR-1360789 and MURI grant 
W911NF-14-1-0003 from ARO. 
A. E. acknowledges financial support by the German National Academy of Sciences 
Leopoldina through grant LPDS~2014-13.
Research at Perimeter Institute is supported by the Government of Canada through Industry Canada 
and by the Province of Ontario through the Ministry of Research and Innovation.

\appendix

\section{One-loop self-energies}
\label{sec:selfen}

The bosonic and fermionic self-energies at one-loop level were already computed 
in Ref.~\cite{Dalidovich2013}. We rederive them here for completeness. The 
following formulas are useful in the derivations. It is often convenient to 
introduce Feynman parameters via
\begin{equation}
	\frac{1}{A^\alpha B^\beta} = 
\frac{\Gamma(\alpha+\beta)}{\Gamma(\alpha)\Gamma(\beta)}\intdefnopi{x}{0}{1} 
\frac{x^{\alpha-1} (1-x)^{\beta-1}}{[x A + (1-x) B]^{\alpha + \beta}}.
\end{equation}
Traces over products of gamma matrices are evaluated using the formulas for 
$2\times2$ matrices, as we are interested in $2 \leq d < 3$,
\begin{gather}
	\tr(\gamma_i) = 0\\
	\tr(\gamma_i \gamma_j) = 2 \delta_{ij}\\
	\tr(\gamma_i \gamma_j \gamma_k \gamma_l) = 2 (\delta_{ij} \delta_{kl} - 
\delta_{ik} \delta_{jl} + \delta_{il} \delta_{jk}),
\end{gather}
where the indices run from $0$ to $d-1$.

At one-loop level, the bosonic self-energy is given by
\begin{equation}
\begin{split}
	\Pi_1(q) &= \frac{e^2 \mu^\epsilon}{N} (d-1) \sum_{j = 1}^N \intd{k}{d+1} 
\tr\bigl(\gamma_1 G_{0,j}(k+q) \gamma_1 G_{0,j}(k)\bigr)\\
	&= -2 e^2 \mu^\epsilon (d-1) \intd{k}{d+1} \frac{\delta_{k+q} \delta_k - \bs 
K 
\cdot (\bs K + \bs Q)}{(\bs K^2 + \delta_k^2) ((\bs K + \bs Q)^2 + 
\delta_{k+q}^2)}
\end{split}
\end{equation}
where $\delta_k = k_x + \sqrt{d-1} k_y^2$. The diagrammatic 
represenation of this contribution is similar to Fig.~\ref{fig:1a}, 
but with current vertices replaced by fermion-boson couplings. Integrating over 
$k_x$, shifting $k_y \rightarrow k_y - \tfrac{\delta_q}{2 q_y}$ and 
integrating over $k_y$ yields
\begin{equation}
	= \frac{e^2 \mu^\epsilon}{4 |q_y|} \sqrt{d-1} \intd{K}{d-1} \Bigl(\frac{\bs K 
\cdot (\bs K + \bs Q)}{|\bs 
K| |\bs K + \bs Q|} - 1 \Bigr),
\end{equation}
where $\int\frac{d^{d-1} K}{(2\pi)^{d-1}} = \int\frac{d K_0}{2\pi} 
\int\frac{d^{1/2-\epsilon} K'}{(2\pi)^{1/2-\epsilon}}$ and $\bs K = K_0 \bs e_0 
+ \bs K'$. The remaining integral can be computed using Feynman parameters, 
yielding
\begin{equation}
	\Pi_1(q) = -\beta_d e^2 \mu^\epsilon \frac{|\bs Q|^{d-1}}{|q_y|},
\end{equation}
where
\begin{displaymath}
	\beta_d = \frac{\sqrt{d-1}\Gamma(d/2)^2}{2^d \sqrt{\pi}^{d-1} |\cos(\frac{\pi 
d}{2})| 
\Gamma(d) \Gamma(\frac{d-1}{2})}.
\end{displaymath}
This result is the same as in Ref.~\onlinecite{Dalidovich2013}.

\begin{figure}
	\centering
	\subfigure[]{\includegraphics[width=2in]{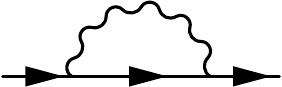} \label{fig3a}}
	\subfigure[]{\includegraphics[width=1.5in]{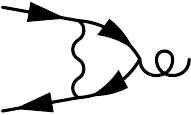} \label{fig3b}}
	\caption{Diagrammatic representation of the one-loop contributions to (a) the 
fermionic self-energy and (b) the current vertex.}
\end{figure}
The fermionic self-energy at one-loop level is given by
\begin{equation}
\begin{split}
	\Sigma_1(q) &= -\frac{e^2 \mu^\epsilon}{N} (d-1) \intd{k}{d+1} D_1(k) 
\gamma_x 
G_0(q-k) \gamma_x\\
	&= \frac{i e^2 \mu^\epsilon}{N} (d-1) \intd{k}{d+1} D_1(k) \frac{\gamma_x 
\delta_{q-k} - \bs \Gamma \cdot (\bs Q - \bs K)}{(\bs Q - \bs K)^2 + 
\delta_{q-k}^2}
\end{split}
\end{equation}
where $(D_1(q))^{-1} = q_y^2 + \beta_d e^2 \mu^\epsilon \frac{|\bs 
Q|^{d-1}}{|q_y|}$ is the one-loop renormalized bosonic propagator. This 
contribution is diagrammatically shown in Fig.~\ref{fig3a}. Shifting $k_x 
\rightarrow k_x + q_x + \sqrt{d-1} (q_y - k_y)^2$, the integrals 
simplify significantly because $\delta_{q-k}$ is effectively replaced by $-k_x$. 
After this shift, $k_y$ appears only in the bosonic propagator, and integration 
over $k_x$ and $k_y$ yields
\begin{equation}
\begin{split}
	\Sigma_1(q) &= -\frac{i e^2 \mu^\epsilon}{N} (d-1) \intd{K}{d-1} 
\intdef{k_y}{-\infty}{\infty} \frac{|k_y|}{|k_y|^3 + \beta_d e^2 \mu^\epsilon 
|\bs K|^{d-1}} \intdef{k_x}{-\infty}{\infty} \frac{\gamma_x k_x + 
\bs \Gamma \cdot (\bs Q - \bs K)}{(\bs Q - \bs K)^2 + k_x^2}\\
&= \frac{i e^{4/3} \mu^{2\epsilon/3} (d-1)}{3 \sqrt{3} \beta_d^{1/3} N} 
\intd{K}{d-1}\frac{\bs \Gamma \cdot (\bs K - \bs Q)}{|\bs 
K|^{\frac{d-1}{3}} |\bs K - \bs Q|}.
\end{split}
\end{equation}
Using Feynman parameters for computing the remaining integral, we obtain 
\begin{equation}
	\Sigma_1(q) = -i (\bs \Gamma \cdot \bs Q) \frac{e^{4/3}}{N} 
\Bigl(\frac{\mu}{|\bs Q|}\Bigr)^{2\epsilon/3} \alpha_{\Sigma,d}
\end{equation}
for the fermionic self-energy at one-loop level in agreement with 
Ref.~\onlinecite{Dalidovich2013}, where
\begin{equation}
\alpha_{\Sigma,d} = \frac{(d-1) \Gamma(\frac{d-1}{3}) \Gamma(\frac{d}{2}) 
\Gamma(\frac{5-2d}{6})}{3 \sqrt{3} \beta_d^{1/3} 2^{d-1} 
\sqrt{\pi}^d \Gamma(\frac{d-1}{6}) \Gamma(\frac{5d-2}{6})}
\end{equation}
For $d = \tfrac{5}{2} - \epsilon$ and $\epsilon \approx 0$, $\alpha_{\Sigma,d}$ 
has a pole in $\epsilon^{-1}$. The pole contribution to the self-energy reads
\begin{equation}
	\Sigma_1(q) = -i(\bs \Gamma \cdot \bs Q) \frac{e^{4/3}}{N} 
\Bigl(\frac{\mu}{|\bs Q|}\Bigr)^{2\epsilon/3} u_{\Sigma,\epsilon} \epsilon^{-1} 
+ \mathcal O(\epsilon^0),
\label{eq:FermionicSelfenergy}
\end{equation}
where $\alpha_{\Sigma,5/2-\epsilon} \approx u_{\Sigma,\epsilon} \epsilon^{-1}$ 
and
\begin{equation}
	u_{\Sigma,\epsilon} = 
\frac{(\frac{3}{2}-\epsilon)\Gamma(\frac{3-2\epsilon}{6}) 
\Gamma(\frac{5-2\epsilon}{4})}{\sqrt{3} \beta_{\frac{5}{2}-\epsilon}^{1/3} 
2^{3/2-\epsilon} \pi^{5/4-\epsilon/2} \Gamma(\frac{3-2\epsilon}{12}) 
\Gamma(\frac{21-10\epsilon}{12})}.
\end{equation}
For $\epsilon = 0$, this reduces to $u_{\Sigma,0} = 
\frac{\Gamma(\frac{5}{4})}{2\sqrt{3}\beta_{5/2}^\frac{1}{3} \pi^\frac{7}{4}}$.

\section{Current vertex at one-loop level}
\label{sec:jj_vertex1Loop}
In this section we derive the one-loop correction to the bare current vertex 
Eq.~\eqref{eq:Gamma0}. It has been computed in Ref.~\onlinecite{Dalidovich2013} 
only for $q = 0$ and here we extend this calculation to $\omega \neq 0$. 

The one-loop correction to the current vertex is shown diagrammatically in 
Fig.~\ref{fig3b} and is given by
\begin{equation}
\begin{split}
	\Gamma_1(k,q) &= -i e_A \frac{e^2 \mu^\epsilon}{N} \intd{p}{d+1} \gamma_x 
G_0(k+p+q) \gamma_x G_0(k+p) \gamma_x D_1(p)\\
	&= \frac{i e_A e^2 \mu^\epsilon}{N} \intd{p}{d+1} D_1(p) \frac{1}{[(\bs K + 
\bs P + \bs Q)^2 + \delta_{k+p+q}^2][(\bs K + \bs P)^2 + \delta_{k+p}^2]}\\
	& \quad \quad \times \gamma_x 
[\bs \Gamma\cdot(\bs K + \bs P + \bs Q) + \gamma_x 
\delta_{k+p+q}]\gamma_x [\bs \Gamma\cdot (\bs K + \bs P) + \gamma_x 
\delta_{k+p}]\gamma_x.
\end{split}
\end{equation}
In the following we set $q = \omega \bs e_0 = \bs Q$. As $\delta_{k+p} = 
k_x + p_x + \sqrt{d-1} (k_y + p_y)^2$, $p_y$ can be eliminated from the 
fermionic propagators by shifting $p_x \rightarrow p_x - k_x - \sqrt{d-1} (k_y + 
p_y)^2$, effectively reducing $\delta_{k+p}$ to $p_x$. Note that $p_y$ still 
appears in the bosonic propagator $D_1$ and that the current vertex correction 
is only a function of $\bs K$. It then reads
\begin{equation}
\begin{split}
	\Gamma_1(\bs K, i \omega) = &i e_A \frac{e^2 \mu^\epsilon}{N} (d-1)
\intd{P}{d-1}\intdef{p_y}{}{}\frac{|p_y|}{|p_y|^3 + \beta_d e^2 \mu^\epsilon}\\
	&\times \intdef{p_x}{}{} \frac{[-\bs \Gamma\cdot(\bs K + \bs P + \bs Q) + 
\gamma_x p_x][\bs \Gamma\cdot(\bs K + \bs P) + \gamma_x 
p_x]\gamma_x}{[(\bs K + \bs P + \bs Q)^2 + p_x^2][(\bs K + \bs P)^2 
+ p_x^2]}.
\label{eq:CurrentVertex_IntermediateResult_1}
\end{split}
\end{equation}
For $\omega = 0$, it is easy to see that the vertex correction vanishes after 
exploiting properties of gamma matrices and computation of the $p_x$ integral. 
Dalidovich and Lee~\cite{Dalidovich2013} argue that this is a sufficient 
condition for the absence of poles in $\epsilon^{-1}$ in $\Gamma_1(\bs K, i 
\omega)$, implying the absence of poles in $\epsilon^{-1}$ in the 
two-loop vertex correction to the current-current correlation function. This is 
checked explicitly below and in Appendix~\ref{app:jj_vertex}.

For an explicit evaluation of the one-loop vertex correction in 
Eq.~\eqref{eq:CurrentVertex_IntermediateResult_1}, we simplify the product in 
the numerator using properties of gamma matrices. All terms that are linear in 
$p_x$ vanish under the integral due to symmetries. Moreover, the $p_y$-integral 
has already been solved during the computation of the fermionic self-energy. 
Rewriting the product in the denominators of the fermionic propgators using 
Feynman parameters, we obtain
\begin{equation}
\begin{split}
\Gamma_1(\bs K, i \omega) = &i e_A \gamma_x \frac{(e^2 
\mu^\epsilon)^{2/3}}{N} \frac{2 (d-1)}{3\sqrt{3} \beta_d^{1/3}} 
\intdefnopi{x}{0}{1} 
\intd{P}{d-1} \intdef{p_x}{}{}\\
	& \times \frac{p_x^2 - \bs \Gamma \cdot (\bs K + \bs P + \bs Q) \Gamma 
\cdot (\bs K + \bs P)}{(\bs P^2)^{\frac{d-1}{6}} \bigl[(\bs P + \bs K + x \bs 
Q)^2 + p_x^2 + x (1-x) \bs Q^2\bigr]^2}.
\end{split}
\end{equation}
In the next step, we perform the $p_x$-integration and write the result in a way 
that makes it transparent that the vertex correction vanishes for $\bs Q = 0$,
\begin{equation}
\begin{split}
= &i e_A \gamma_x \frac{(e^2 \mu^\epsilon)^{2/3} (d-1)}{6 \sqrt{3} N 
\beta_d^{1/3}} \intdefnopi{x}{0}{1} \intd{P}{d-1} \\
	& \times \frac{(\bs P + \bs K + x \bs Q)^2 + x(1-x) \bs Q^2 - \bs \Gamma 
\cdot (\bs K + \bs P + \bs Q) \bs \Gamma \cdot (\bs K + \bs P)}{(\bs 
P^2)^\frac{d-1}{6} \bigl[(\bs P + \bs K + x \bs Q)^2 + x(1-x) \bs 
Q^2\bigr]^{3/2}}.
\end{split}
\end{equation}
The product in the denominator can further be rewritten using Feynman 
parameters, yielding
\begin{equation}
\begin{split}
= &i e_A \gamma_x \frac{(e^2 \mu^\epsilon)^{2/3}}{6 \sqrt{3} N 
\beta_d^{1/3}} 
\frac{\Gamma(\frac{3}{2} + 
\frac{d-1}{6}) (d-1)}{\Gamma(\frac{3}{2}) 
\Gamma(\frac{d-1}{6})}\intdefnopi{x}{0}{1}\intdefnopi{y}{0}{1}\intd{P}{d-1} 
y^{\frac{d-1}{6}-1} (1-y)^{1/2} \\
	&\times \frac{(\bs P + \bs K + x \bs Q)^2 + x (1-x) \bs Q^2 - \bs\Gamma \cdot 
(\bs K + \bs P + \bs Q) \bs\Gamma \cdot (\bs K + \bs P)}{\bigl[(\bs P + 
(1-y)(\bs K + x \bs Q))^2 + y(1-y) (\bs K + x \bs Q)^2 + x(1-x) (1-y) \bs 
Q\bigr]^{\frac{3}{2} + \frac{d-1}{6}}}
\end{split}
\end{equation}
after completing squares in the denominator. We next shift $\bs P \rightarrow 
\bs P - (1-y)(\bs K + x \bs Q)$. Terms in the numerator that are odd in $\bs P$ 
vanish when computing the integral, and we obtain
\begin{equation}
\begin{split}
\Gamma_1(\bs K, i\omega) &= i e_A \gamma_x \tilde\Gamma_1(\bs K, i\omega)\\
	&= i e_A \gamma_x \frac{(e^2 \mu^\epsilon)^{2/3}}{6 \sqrt{3} N 
\beta_d^{1/3}} 
\frac{\Gamma(\frac{3}{2} + 
\frac{d-1}{6}) (d-1)}{\Gamma(\frac{3}{2}) 
\Gamma(\frac{d-1}{6})}\intdefnopi{x}{0}{1}\intdefnopi{y}{0}{1}\intd{P}{d-1} 
y^\frac{d-7}{6} (1-y)^{1/2} \\
	&\qquad \times \frac{x \bs Q \cdot \bigl[(1-2x(1-y)) \bs Q + 2y 
\bs K\bigr] - \bs \Gamma \cdot \bs Q \bs \Gamma\cdot (y \bs K - x(1-y) 
\bs Q)}{\bigl[\bs P^2 + y (1-y) (\bs K + x \bs Q)^2 + 
x(1-x)(1-y) \bs Q^2\bigr]^{\frac{3}{2} + \frac{d-1}{6}}}
\label{eq:OneLoopCurrentVertex}
\end{split}
\end{equation}
with $\tilde \Gamma_1(\bs K, i\omega)$ defined in an obvious way. This result 
is used in Appendix~\ref{app:jj_vertex} for the computation of the two-loop 
vertex correction to the current-current correlation function.

\section{Evaluation of current-current correlation function}

\subsection{Free fermion contribution}
\label{app:jj_free}
The free fermion contribution to the current-current correlation function in 
Eq.~\eqref{eq:jj_1Loop} is straightforwardly evaluated using dimensional 
regularization. Shifting $k_x \rightarrow k_x - \sqrt{d-1} k_y^2$, $k_y$ 
disappears completely from the integrand, yielding
\begin{equation}
\begin{split}
\langle J_x J_x\rangle_\text{1Loop}(i\omega) &= -2 e^2_A N 
\intdef{k_y}{}{}\intdef{k_x}{}{} 
\intd{K}{d-1} \frac{k_x^2 - \bs K 
\cdot (\bs K + \bs Q)}{(\bs K^2 + k_x^2) ((\bs K + \bs Q)^2 + k_x^2)} \\
	&= -2 e^2_A N \intdef{k_y}{}{} I_\text{1loop}(\bs Q).
\end{split}
\end{equation}
Introducing Feynman parameters, completing squares in the denominator and 
shifting $\bs K \rightarrow \bs K - (1-x) \bs Q$, we obtain
\begin{equation}
\begin{split}
	I_\text{1loop}(\bs Q) &= \intd{K}{d-1}\intd{p}{} \intdefnopi{x}{0}{1} 
\frac{p^2 - \bs K^2 + x (1-x) 
\bs Q^2}{[\bs K^2 + p^2 + x (1 - x) \bs Q^2]^2}\\
&= \frac{\pi S_{d-1}}{(2\pi)^d} \intdefnopi{k}{0}{\infty} k^{d-2} 
\intdefnopi{x}{0}{1} \frac{ x (1-x) \bs Q^2}{[k^2 + x (1-x) \bs Q^2]^{3/2}}\\
&= \frac{S_{d-1}}{(2\pi)^d} \sqrt{\pi} \Gamma(2 - d/2) 
 \frac{\Gamma(\frac{d-1}{2})\Gamma(d/2)^2}{\Gamma(d)} |\bs Q|^{d-2}.
\end{split}
\end{equation}
For $d = 5/2 - \epsilon$, the one-loop result for the current-current 
correlation function thus reads
\begin{equation}
	\langle J_x J_x \rangle_\text{1loop} (i\omega) = -e^2_A N 
u_{\text{1Loop},\epsilon}
\intdef{k_y}{}{} |\omega|^{1/2-\epsilon},
\end{equation}
where
\begin{equation}
u_{\text{1Loop},\epsilon} = \frac{2^{\epsilon-1/2} 
\Gamma(\frac{3+2\epsilon}{4}) 
\Gamma(\frac{5-2\epsilon}{4})^2}{\sqrt{\pi}^{5/2-\epsilon} 
\Gamma(\frac{5-2\epsilon}{2})}.
\end{equation}

\subsection{Self-energy correction}
\label{app:jj_selfenergy}
The two-loop self-energy correction to the current-current correlation 
function, Eq.~\eqref{eq:jj_selfenergy1}, can be computed using Feynman 
parameters. After rewriting the integrand it reads
\begin{equation}
	\langle J_x J_x\rangle_\text{SE}(\omega) =4 (e^2\mu^\epsilon)^\frac{2}{3} 
e_A^2 \alpha_{\Sigma,d} \intd{k}{d+1} 
\intdefnopi{x}{0}{1} \frac{1-x}{|\bs K|^\frac{2\epsilon}{3}} \frac{2\delta_k^2 
\bs K^2 + \bs K \cdot (\bs K + \bs Q)(\delta_k^2 - \bs K^2)}{\bigl[x (\bs K + 
\bs Q)^2 + (1-x) \bs K^2 + \delta_k^2\bigr]^3}.
\end{equation}
Eliminating $k_y$ by a variable shift of $k_x$ and subsequent integration over 
$k_x$ yield
\begin{equation}
\begin{split}
	=\frac{\Gamma(3)}{4}(e^2\mu^\epsilon)^\frac{2}{3} e_A^2 \alpha_{\Sigma,d} 
&\intdef{k_y}{}{} \intd{K}{d-1} \intdefnopi{x}{0}{1} \frac{1-x}{|\bs 
K|^\frac{2\epsilon}{3}} \\
	& \times \Bigl[\frac{3\bs K^2 + \bs K \cdot \bs Q}{\bigl[\bs 
K^2 + x (2\bs K \cdot \bs Q + \bs Q^2)\bigr]^\frac{3}{2}} - \frac{3 \bs K^2 
(\bs K^2 + \bs K \cdot \bs Q)}{\bigl[\bs K^2 + x (2 \bs K \cdot \bs Q + \bs 
Q^2)\bigr]^\frac{5}{2}}\Bigr].
\end{split}
\end{equation}
Again using Feynman parameters to rewrite the products in the integrand, we 
obtain
\begin{equation}
\begin{split}
	=\frac{\Gamma(3)}{4\Gamma(\frac{\epsilon}{3})}& e_A^2 
(e^2\mu^\epsilon)^\frac{2}{3} \alpha_{\Sigma,d} 
\intdef{k_y}{}{}\intd{K}{d-1}\intdefnopi{x}{0}{1}\intdefnopi{y}{0}{1} \\
&\times \Bigl[\frac{\Gamma(\frac{9+2\epsilon}{6})}{\Gamma(\frac{3}{2})} 
\frac{(1-x) y^{\frac{\epsilon}{3}-1} (1-y)^\frac{1}{2} (3 \bs K^2 + \bs K \cdot 
\bs Q)}{\bigl[\bs K^2 + x (1-y) (2\bs K \cdot \bs Q + \bs 
Q^2)]^{\frac{3}{2}+\frac{\epsilon}{3}}} \\
	&\quad - 
\frac{\Gamma(\frac{15+2\epsilon}{6})}{\Gamma(\frac{5}{2})} \frac{3 (1-x) 
y^{\frac{\epsilon}{3}-1} (1-y)^\frac{3}{2} \bs K^2 (\bs K^2 + \bs K \cdot \bs 
Q)}{\bigl[\bs K^2 + x (1-y) (2 \bs K \cdot \bs Q + \bs Q^2)\bigr]^{\frac{5}{2} 
+ \frac{\epsilon}{3}}}\Bigr].
\end{split}
\end{equation}
Completing squares in the denominator as
\begin{equation}
	\bs K^2 + x (1-y)(2 \bs K \cdot \bs Q + \bs Q^2) = (\bs K + x(1-y)\bs Q)^2 + 
x (1-y) (1-x + xy) \bs Q^2,
\end{equation}
 shifting $\bs K \rightarrow \bs K - x (1-y) \bs Q$, and neglecting terms that 
vanish due to symmetries when performing the $\bs K$-integration, we obtain
\begin{equation}
\begin{split}
	=&\frac{\Gamma(3)}{4\Gamma(\frac{\epsilon}{3})} e_A^2 (e^2 
\mu^\epsilon)^\frac{2}{3} \alpha_{\Sigma,d} \intd{k_y}{} \intd{K}{d-1} 
\intdefnopi{x}{0}{1} \intdefnopi{y}{0}{1} (1-x) 	y^{\frac{\epsilon}{3}-1}\\
	& \times \Bigl\{\frac{\Gamma(\frac{9+2\epsilon}{6})}{\Gamma(\frac{3}{2})} 
(1-y)^\frac{1}{2} \frac{3 \bs K^2 - x (1-y) (1-3x(1-y)) \bs Q^2}{\bigl[\bs K^2 
+ 
x (1-y) (1-y+xy) \bs Q^2\bigr]^{\frac{3}{2}+\frac{\epsilon}{3}}} \\
	&\quad - \frac{\Gamma(\frac{15+2\epsilon}{6})}{\Gamma(\frac{5}{2})} 
\frac{3 (1-y)^\frac{3}{2}}{\bigl[\bs K^2 + 
x(1-y) (1-x+xy) \bs Q^2\bigr]^{\frac{5}{2} + \frac{\epsilon}{3}}} 
\Bigl[\bs K^4 - x(1-y)(1-2x(1-y)) \bs K^2 \bs Q^2 \\
	& \quad \quad \quad - 2 x (1-y) (1-2x 
(1-y)) (\bs K \cdot \bs Q)^2 - x^3 (1-y)^3 (1-x(1-y)) \bs Q^4\Bigr]\Bigr\}
\end{split}
\end{equation}
The remaining integrals can easily be computed using \verb|Mathematica|. First 
integrating over $\bs K$ and subsequently over $x$ and $y$, the pole 
contribution to the two-loop self-energy correction reads
\begin{equation}
	\langle J_x J_x \rangle_\text{SE}(i\omega) = \frac{\pi^\frac{1}{4} 
u_{\Sigma,\epsilon = 0}}{8 \sqrt{2} \Gamma(\frac{7}{4})} e_A^2 e^{4/3} 
\epsilon^{-1} \intd{k_y}{} |\omega|^{\frac{1}{2} - \epsilon} 
\Bigl(\frac{\mu}{|\omega|}\Bigr)^{2\epsilon/3} + \mathcal O(\epsilon^0)
\end{equation}
after exploiting $\alpha_{\Sigma,d} \approx u_{\Sigma,\epsilon} \epsilon^{-1}$ 
for $\epsilon \approx 0$ and setting $\epsilon$ to zero in the numerical 
prefactors. 

\subsection{Vertex correction contribution}
\label{app:jj_vertex}
In the following, we briefly describe the evaluation of the two-loop vertex 
correction contribution to the current-current correlation function, 
Eq.~\eqref{eq:jj_2Loop_VC}. Eliminating $k_y$ from the integrand by shifting 
$k_x \rightarrow k_x - \sqrt{d-1} k_y^2$ and expanding the products of gamma 
matrices in the numerator, the integrand simplifies because all terms in the 
numerator that are odd in $k_x$ vanish due to symmetries. We obtain
\begin{equation}
	\langle J_x J_x \rangle_\text{VC} (i\omega) = -e_A^2 N \intd{k}{d+1} 
\operatorname{tr}\Bigl[\frac{k_x^2 - \bs 
\Gamma\cdot \bs K \bs \Gamma \cdot (\bs K + \bs Q)}{(\bs K^2 + k_x^2)((\bs 
K + \bs Q)^2 + k_x^2)} \tilde \Gamma_1(\bs K, i\omega)\Bigr].
\end{equation}
Introduction of Feynman parameters and subsequent integration over $k_x$ 
yields
\begin{equation}
\begin{split}
	=-e_A^2 N &\intd{K}{d-1} \intdef{k_y}{}{} \intdefnopi{z}{0}{1}\\
	& \times\operatorname{tr} \Bigl[\frac{\bs K^2 + (1-z) (2 \bs K \cdot \bs Q + 
\bs Q^2) - \bs \Gamma \cdot \bs K \bs \Gamma \cdot (\bs K + \bs Q)}{4\bigl[\bs 
K^2 + (1-z)(2 \bs K \cdot \bs Q + \bs Q^2)\bigr]^{3/2}} \tilde \Gamma_1(\bs K, 
i\omega)\Bigr].
\end{split}
\end{equation}

Inserting the one-loop correction to the current vertex in 
Eq.~\eqref{eq:OneLoopCurrentVertex}, we obtain
\begin{equation}
\begin{split}
&=-e_A^2 \frac{(e^2 \mu^\epsilon)^{2/3}}{24 \sqrt{3} \beta_d^{1/3}} 
\frac{\Gamma(\frac{3}{2} + 
\frac{d-1}{6}) (d-1)}{\Gamma(\frac{3}{2})\Gamma(\frac{d-1}{6})} 
\intdefnopi{x}{0}{1} 
\intdefnopi{y}{0}{1} \intdefnopi{z}{0}{1} \intd{K}{d-1} 
\intd{P}{d-1}\intdef{k_y}{}{} y^\frac{d-7}{6} (1-y)^{1/2}\\
&\qquad \times \operatorname{tr}\Bigl[\frac{\bs K^2 + (1-z)(2 \bs K \cdot \bs Q 
+ \bs Q^2) - \bs \Gamma \cdot \bs K \bs \Gamma \cdot (\bs K + \bs Q)}{\bigl[\bs 
K^2 + (1-z)(2 \bs K \cdot \bs Q + \bs Q^2)\bigr]^{3/2}}\\
&\qquad\quad \times\frac{x \bs Q \cdot \bigl[(1-2x(1-y))\bs Q + 2 y \bs K\bigr] 
- \bs \Gamma \cdot \bs Q \bs \Gamma \cdot (y \bs K - x(1-y)\bs Q)}{\bigl[\bs 
P^2 + y(1-y)(\bs K + x \bs Q)^2 + x(1-x)(1-y) \bs Q^2\bigr]^{\frac{3}{2} + 
\frac{d-1}{6}}}\Bigr].
\end{split}
\end{equation}
Again introducing Feynman parameters for the remaining product yields
\begin{equation}
\begin{split}
&= -e_A^2 \frac{(e^2 \mu^\epsilon)^{2/3}}{24\sqrt{3} \beta_d^{1/3}} 
\frac{\Gamma(3 + \frac{d-1}{6}) (d-1)}{\Gamma(\frac{3}{2})^2 
\Gamma(\frac{d-1}{6})} 
\intdefnopi{w}{0}{1} \intdefnopi{x}{0}{1} \intdefnopi{y}{0}{1} 
\intdefnopi{z}{0}{1} \intd{K}{d-1} \intd{P}{d-1} \intdef{k_y}{0}{1} \\
&\quad \times y^\frac{d-7}{6} (1-y)^{1/2}\operatorname{tr} \Bigl(\bigl\{\bs K^2 
+ (1-z) (2 \bs K \cdot \bs Q + \bs Q^2) - \bs \Gamma \cdot \bs K \bs \Gamma 
\cdot (\bs K + \bs Q)\bigr\}\\
&\quad \quad \quad \quad  \times \bigl\{x \bs Q\cdot [(1-2(1-y))\bs Q + 2y\bs 
K] - \bs \Gamma \cdot \bs Q \bs \Gamma \cdot (y \bs K - x(1-y) \bs 
Q)\bigr\}\Bigr)\\
	&\quad \times w^{1/2} (1-w)^\frac{d+2}{6} \Bigl[(1-w) \bs P^2 + 
[w+(1-w)y(1-y)](\bs K + \alpha_1 \bs Q)^2 + \alpha_2^2 \bs Q^2\Bigr]^{-(3 + 
\frac{d-1}{6})}\\
\end{split}
\end{equation}
after completion of squares in the denominator and definition of
\begin{gather}
	\alpha_1(w,x,y,z) = \alpha_1 = \frac{(1-w)x y (1-y) + w (1-z)}{w + 
(1-w)y(1-y)}\\
\begin{split}
	\alpha_2(w,x,y,z) = & \alpha_2 = \bigl[w(1-z) + x(1-x)(1-w)(1-z) + (1-w)x^2 y 
(1-y) \\
		&\quad - \alpha_1^2 (w + (1-w)y(1-y))\bigr]^{1/2}.
\end{split}
\end{gather}
In the next step, we shift $\bs K \rightarrow \bs K - \alpha_1 \bs Q$ and 
subsequently evaluate the trace over gamma matrices. Terms in the numerator 
which are odd in $\bs K$ vanish under the integral due to symmetries. The trace 
over gamma matrices then yields
\begin{equation}
\begin{split}
	\operatorname{tr}(\ldots) &= 2y \bs K^2 \bs Q^2 - 4y(1-x-z+2xz) (\bs K \cdot 
\bs Q)^2 \\
&\quad + 2 \bs Q^4 \bigl(1-\alpha_1 - z(1-2\alpha_1)\bigr) \bigl(y \alpha_1 
- 2 x^2 (1-y) + x(2-y -2 y \alpha_1)\bigr).
\end{split}
\end{equation}
No contribution $\sim \bs K^4$ exists because the vertex correction vanishes for 
$|\bs Q| = |\omega| \rightarrow 0$.

Rescaling $\bs K$ and $\bs P$ as
\begin{align*}
	\bs P &\rightarrow \frac{\alpha_2}{\sqrt{1-w}} \bs P		&		\bs K 
&\rightarrow \frac{\alpha_2}{\sqrt{w + (1-w)y (1-y)}} \bs K,
\end{align*}
we obtain
\begin{equation}
\begin{split}
\langle J_x J_x\rangle_\text{VC}(i\omega) &= -e_A^2 \frac{(e^2 
\mu^\epsilon)^{2/3}}{24 \sqrt{3} \beta_d^{1/3}} \frac{\Gamma(3 + 
\frac{d-1}{6}) (d-1)}{\Gamma(\frac{3}{2})^2 \Gamma(\frac{d-1}{6})} 
\intdef{k_y}{}{} \\
	& \times \Bigl( N_\text{VC}^{(1)} \bs Q^2 S_\text{VC}^{(1)}(\bs Q) + 
N_\text{VC}^{(2)} S_\text{VC}^{(2)}(\bs Q) + N_\text{VC}^{(3)} \bs Q^4 
S_\text{VC}^{(3)}(\bs Q)\Bigr)
\end{split}
\end{equation}
where
\begin{equation}
\begin{split}
	S_\text{VC}^{(1)}(\bs Q) &= \intd{K}{d-1}\intd{P}{d-1} \frac{\bs 
K^2}{\bigl[\bs P^2 + \bs K^2 + \bs Q^2\bigr]^{3 + \frac{d-1}{6}}} = 
\frac{(3-2\epsilon) \Gamma(\frac{3}{4} + 
\frac{5\epsilon}{6})}{\Gamma(\frac{13}{4} - \frac{\epsilon}{6}) 2^{5-2\epsilon} 
\pi^{3/2-\epsilon}} (\bs Q^2)^{-\frac{3}{4} - \frac{5\epsilon}{6}}
\end{split}
\end{equation}
\begin{equation}
\begin{split}
	S_\text{VC}^{(2)}(\bs Q) &= \intd{K}{d-1} \intd{P}{d-1} \frac{(\bs K \cdot 
\bs Q)^2}{\bigl[\bs P^2 + \bs K^2 + \bs Q^2\bigr]^{3 + \frac{d-1}{6}}} = 
\frac{\Gamma(\frac{3}{4} + \frac{5\epsilon}{6})}{4^{2-\epsilon} \pi^{3/2 - 
\epsilon} \Gamma(\frac{13}{4} - \frac{\epsilon}{6})} (\bs Q^2)^{\frac{1}{4} - 
\frac{5\epsilon}{6}}
\end{split}
\end{equation}
\begin{equation}
\begin{split}
 S_\text{VC}^{(3)}(\bs Q) &= \intd{K}{d-1}\intd{P}{d-1} \frac{1}{\bigl[\bs P^2 
+ \bs K^2 + \bs Q^2\bigr]^{3 + \frac{d-1}{6}}} = \frac{\Gamma(\frac{7}{4} + 
\frac{5\epsilon}{6})}{2^{3-2\epsilon} \pi^{3/2 - \epsilon} \Gamma(\frac{13}{4} 
- 
\frac{\epsilon}{6})} (\bs Q^2)^{-\frac{7}{4} - \frac{5\epsilon}{6}}
\end{split}
\end{equation}
\begin{equation}
\begin{split}
N_\text{VC}^{(1)} &= \intdefnopi{w}{0}{1} \intdefnopi{x}{0}{1} 
\intdefnopi{y}{0}{1} \intdefnopi{z}{0}{1} 
\Bigl(\frac{\alpha_2^2}{\sqrt{w+(1-w)y(1-y)}\sqrt{1-w}}\Bigr)^{d-1} 
y^\frac{d-7}{6} (1-y)^{1/2} w^{1/2} \\
&\quad \times (1-w)^\frac{d+2}{6} (\alpha_2^2)^{-(3 + 
\frac{d-1}{6})} \frac{2 y \alpha_2^2}{w + (1-w) y (1-y)}
\end{split}
\end{equation}
\begin{equation}
\begin{split}
N_\text{VC}^{(2)} &= \intdefnopi{w}{0}{1} \intdefnopi{x}{0}{1} 
\intdefnopi{y}{0}{1} \intdefnopi{z}{0}{1} 
\Bigl(\frac{\alpha_2^2}{\sqrt{w+(1-w)y(1-y)}\sqrt{1-w}}\Bigr)^{d-1} 
y^\frac{d-7}{6} (1-y)^{1/2} w^{1/2} \\
&\quad \times (1-w)^\frac{d+2}{6} (\alpha_2^2)^{-(3 + 
\frac{d-1}{6})} \frac{-4 y \alpha_2^2 (1-x-z+2xz)}{w + (1-w)y(1-y)}
\end{split}
\end{equation}
\begin{equation}
\begin{split}
N_\text{VC}^{(3)} &= \intdefnopi{w}{0}{1} \intdefnopi{x}{0}{1} 
\intdefnopi{y}{0}{1} \intdefnopi{z}{0}{1} 
\Bigl(\frac{\alpha_2^2}{\sqrt{w+(1-w)y(1-y)}\sqrt{1-w}}\Bigr)^{d-1} 
y^\frac{d-7}{6} (1-y)^{1/2} w^{1/2} \\
&\quad \times (1-w)^\frac{d+2}{6} (\alpha_2^2)^{-(3 + 
\frac{d-1}{6})} 2 (1-\alpha_1 - z(1-2\alpha_1)) \\
	&\quad \times (y \alpha_1 - 2 x^2(1-y) + x 
(2-y - 2 y \alpha_1)).
\end{split}
\end{equation}
These integrals can easily be computed using \verb|Mathematica|. They are free 
of poles in $\epsilon^{-1}$, so that $\epsilon$ can be set to zero in
numerical prefactors. This yields
\begin{equation}
\langle J_x J_x\rangle_\text{VC} (i\omega) = -\alpha_\text{VC}^{\epsilon = 0} 
e_A^2 e^{4/3} |\omega|^{\frac{1}{2} - \epsilon} 
\Bigl(\frac{\mu}{|\omega|}\Bigr)^{2\epsilon/3} \intdef{k_y}{}{}
\end{equation}
where $\alpha_\text{VC}^{\epsilon = 0} \approx 0.0230903$.

\section{Particle current-momentum susceptibility}
\label{sec:JCPsusceptibility}
Following the arguments in Sec.~\ref{sec:current-momentum}, we expect that the 
particle current-momentum susceptibility is non-zero, at least in $d = 2$. It 
is given by
\begin{equation}
	\chi_{J^N, P} = \lim_{\bs q \rightarrow 0} \langle J^N_x 
P_x\rangle (q_0 = 0, \bs q),
\end{equation}
where $J_x^N$ is the $x$-component of the particle current. The correlation 
function for $\bs q \neq \bs 0$ reads
\begin{equation}
\begin{split}
	\langle J_x^N P_x&\rangle_\text{1Loop} (q_0 = 0, \bs q) = -e_N N 
\intd{k}{d+1} \bigl(k_x + \frac{q_x}{2}\bigr) 
\operatorname{tr}\bigl(\gamma_0 G_0(k+q) \gamma_0 
G_0(k)\bigr)\\
	&= e_N N 
\intd{k}{d+1} \bigl(k_x + \frac{q_x}{2}\bigr) 
\frac{\operatorname{tr}\bigl\{ \gamma_0 [\bs \Gamma\cdot \bs K + \gamma_x 
\delta_{k+q}]\gamma_0 [\bs \Gamma\cdot \bs K + \gamma_x 
\delta_k]\bigr\}}{(\bs K^2 + \delta_{k+q}^2) (\bs K^2 + \delta_k^2)}.
\label{eq:DensCurrentMomSusceptibility}
\end{split}
\end{equation}
The trace over the gamma matrices yields
\begin{equation}
	= 2 e_N N 
\intd{k}{d+1} \bigl(k_x + \frac{q_x}{2}\bigr)  \frac{2 K_0^2 - \bs K^2 - 
\delta_k \delta_{k+q}}{(\bs K^2 + \delta_{k+q}^2) (\bs K^2 + \delta_k^2)}.
\end{equation}
It is advantageous to split the physical from the auxiliary frequency 
directions,
\begin{equation}
	\bs K = K_0 \bs e_0 + \bs K',
\end{equation}
in terms of which we obtain
\begin{gather}
	= 2 e_N N \intdefnopi{x}{0}{1} \intd{k}{d+1} \bigl(k_x + \frac{q_x}{2}\bigr)  
\frac{K_0^2 - \bs K'^2 - \delta_k \delta_{k+q}}{(K_0^2 + \bs K'^2 + x \delta_k^2 
+ (1-x)\delta_{k+q}^2)^2}
\end{gather}
after introducing Feynman parameters. Shifting $k_x \rightarrow k_x - 
\sqrt{d-1} k_y^2$, introducing $G = q_x + \sqrt{d-1} (2 k_y q_y + q_y^2)$ and 
completing squares in the denominator yields
\begin{equation}
	= 2 e_N N \intdefnopi{x}{0}{1} \intd{k}{d+1} \frac{\bigl(k_x - \sqrt{d-1} 
k_y^2 + \frac{q_x}{2}\bigr) \bigl(K_0^2 - \bs K'^2 - k_x (k_x + 
G)\bigr)}{\bigl[K_0^2 + \bs K'^2 + (k_x + (1-x)G)^2 + x(1-x) G^2\bigr]^2}.
\end{equation}
After shifting $k_x \rightarrow k_x - (1-x) G$, all terms in the numerator 
which are odd in $k_x$ vanish and we obtain
\begin{equation}
\begin{split}
	= 2 e_N N \intdefnopi{x}{0}{1}\intd{k}{d+1} & 
\Bigr\{(1-2x) G \frac{k_x^2}{[K_0^2 + \bs K'^2 + k_x^2 + x(1-x) G^2]^2} 
\\ & - \bigl[\sqrt{d-1} k_y^2 - \frac{q_x}{2} + (1-x) G\bigr] \frac{K_0^2 - \bs 
K'^2 - 
k_x^2 + x(1-x) G^2}{[K_0^2 + \bs K'^2 + k_x^2 + x(1-x) G^2]^2}\Bigr\}.
\end{split}
\end{equation}
It is obvious that the contribution in the first line vanishes when performing 
the $x$-integration. Rescaling integration variables as $k_0 \rightarrow 
\sqrt{x(1-x)} k_0$, $k_x \rightarrow \sqrt{x(1-x)} k_x$ and $\bs K' \rightarrow 
\sqrt{x(1-x)} \bs K'$, the term in the second line reads
\begin{equation}
\begin{split}
	= -2 e_N N& \intd{k_y}{} \intd{k_0}{} \intd{k_x}{} 
\intd{K'}{\frac{1}{2}-\epsilon} 
\frac{K_0^2 - \bs K'^2 - k_x^2 + G^2}{[K_0^2 + \bs K'^2 + k_x^2 + 
G^2]^2}\\
&\times \intdefnopi{x}{0}{1} \sqrt{x(1-x)}^{d-2} \bigl[\sqrt{d-1} k_y^2 - 
\frac{q_x}{2} + (1-x)G\bigr]
\end{split}
\end{equation}
The remaining integrals yield
\begin{gather}
	\intdefnopi{x}{0}{1} \sqrt{x(1-x)}^{d-2} \bigl[\sqrt{d-1} k_y^2 - 
\frac{q_x}{2} + 
(1-x) G\bigr] = 
\frac{\Gamma(1+\frac{d}{2}) \Gamma(\frac{d}{2})}{\Gamma(1+d)} G + 
\frac{\Gamma(\frac{d}{2})^2}{\Gamma(d)} \bigl(\sqrt{d-1} k_y^2 - 
\frac{q_x}{2}\bigr)\\
\begin{split}
 \intd{k_0}{} \intd{k_x}{} \intd{K'}{\frac{1}{2}-\epsilon}& \frac{K_0^2 - \bs 
K'^2 - 
k_x^2 + G^2}{[k_0^2 + \bs K'^2 + k_x^2 + G^2]^2} = \intd{k_x}{} 
\intd{K'}{\frac{1}{2}-\epsilon} \frac{G^2}{2\sqrt{G^2 + k_x^2 + \bs K'^2}^3}\\
&= \frac{1}{2\pi} \intd{K'}{\frac{1}{2}-\epsilon} \frac{G^2}{G^2 + \bs K'^2} = 
-\frac{|G|^{d-2}}{2^{d-1} \pi^{\frac{d}{2} - 1} \sin(\frac{\pi d}{2}) 
\Gamma(\frac{d-2}{2})}.
\end{split}
\end{gather}
We thus obtain
\begin{equation}
\begin{split}
	\langle J_x^N P_x \rangle_\text{1Loop} (q_0 = 0, \bs q) = e_N N 
& \frac{\sqrt{d-1}\Gamma(\frac{d}{2})^2}{\pi^{\frac{d}{2}-1} \sin(\frac{\pi 
d}{2}) \Gamma(\frac{d-2}{2}) \Gamma(d)} \intd{k_y}{} \bigl(k_y q_y + 
\frac{q_y^2}{2} + k_y^2\bigr) \\
	&\times \Bigl|\frac{q_x}{2} + \sqrt{d-1} \Bigl(k_y q_y + 
\frac{q_y^2}{2}\Bigr)\Bigr|^{d-2}
\end{split}
\label{eq:chi_JN_P}
\end{equation}
for the particle current-momentum correlation function. Note that on one-loop 
level the limits $d\rightarrow 2$ and $\bs q \rightarrow \bs 0$ do not commute. 
For $d \geq 2$, we obtain
\begin{gather}
	\lim_{d\rightarrow 2} \lim_{\bs q \rightarrow \bs 0} \langle J_x^N 
P_x\rangle_\text{1Loop}(0, 0, q_y) = 0\\
\lim_{\bs q \rightarrow \bs 0} \lim_{d\rightarrow 2} \langle J_x^N 
P_x\rangle_\text{1Loop}(0, 0, q_y) = -\frac{e_N N}{\pi} \intd{k_y}{} k_y^2.
\end{gather}
The result in the last line also follows from a calculation in $d = 2$.

\section{One-loop conductivity at finite temperature}
\label{sec:sigmaTg0}

In this section, we compute the one-loop result for the conductivity in the 
limit $\omega \ll T$ for $d = 5/2 - \epsilon$. It is given by
\begin{equation}
\begin{split}
	\sigma^\text{1Loop}_{xx}(i\Omega_m) &= -\frac{1}{\Omega_m} \langle 
J_x J_x \rangle_\text{1Loop} (i\Omega_m)\\
 &= \frac{2 e_A^2 N}{\beta \Omega_m} \intd{k}{2} \intd{K'}{\frac{1}{2}-\epsilon} 
\sum_n 
\frac{\delta_k^2 - \bs K'^2 - \omega_n(\omega_n + \Omega_m)}{(\omega_n^2 + \bs 
K'^2 + \delta_k^2)\bigl((\omega_n + \Omega_m)^2 + \bs K'^2 + \delta_k^2\bigr)},
\end{split}
\end{equation}
where $\Omega_m$ is a bosonic Matsubara frequency that has to be analytically 
continued to real frequencies, $i \Omega_m \rightarrow \omega + i \eta$, 
after evaluation of the sum over fermionic Matsubara frequencies $\omega_n$. 

Before summing over fermionic Matsubara frequencies, we cast this equation in a 
form that makes it explicit that $\Omega_m \sigma^\text{1Loop}_{xx}(i\Omega_m)$ 
vanishes for $\Omega_m = 0$, following Ref.~\onlinecite{Sachdev1998},
\begin{equation}
\begin{split}
	&= \frac{e_A^2 N}{\beta \Omega_m} \sum_{\omega_n} \intd{k}{2} 
\intd{K'}{\frac{1}{2}-\epsilon} 
\Bigl\{\frac{\Omega_m^2 + 4 k_x^2}{\bigl(\omega_n^2 + \bs K'^2 + 
k_x^2\bigr)\bigl[(\omega_n + \Omega_m)^2 + \bs K'^2 + k_x^2\bigr]} - 
\frac{2}{\omega_n^2 + \bs K'^2 + k_x^2}\Bigr\}\\
	&= \frac{e_A^2 N}{\beta \Omega_m} \sum_{\omega_n} \intd{k}{2} 
\intd{K'}{\frac{1}{2}-\epsilon} 
\frac{1}{\omega_n^2 + \bs K'^2 + k_x^2} \Bigl\{\frac{\Omega_m^2 + 4 
k_x^2}{(\omega_n + \Omega_m)^2 + \bs K'^2 + k_x^2} - \frac{4 
k_x^2}{\omega_n^2 + \bs K'^2 + k_x^2}\Bigr\}.
\end{split}
\end{equation}

Summing over the fermionic Matsubara frequencies yields
\begin{equation}
\begin{split}
	&= \frac{e_A^2 N}{\Omega_m} \intd{k}{2} \intd{K'}{\frac{1}{2}-\epsilon} 
\Bigl\{\frac{\Omega_m^2 + 4 k_x^2}{\Delta_k (\Omega_m^2 + 4 \Delta_k^2)} 
\bigl[ n_F(-\Delta_k) - n_F(\Delta_k)\bigr] \\
	&\qquad \qquad - \frac{k_x^2}{\Delta_k^3} 
\bigl[n_F(-\Delta_k) - n_F(\Delta_k) + \Delta_k \bigl(n_F'(\Delta_k) + 
n_F'(-\Delta_k)\bigr)\bigr]\Bigr\},
\end{split}
\end{equation}
where $\Delta_k = \sqrt{k_x^2 + \bs K'^2}$. Analytical continuation using 
$i \Omega_m \rightarrow \omega + i \delta$ with $\delta = 0^+$ yields
\begin{equation}
\begin{split}
	\sigma_{xx}^\text{1Loop}(\omega,T) &= \frac{ie_A^2 N}{\omega+i\delta} 
\intd{k}{2} \intd{K'}{\frac{1}{2}-\epsilon} 
\Bigl\{\frac{(-i\omega+\delta)^2 + 4 k_x^2}{\Delta_k 
\bigl[(-i\omega+\delta)^2 + 4 \Delta_k^2\bigr]} 
\bigl[ n_F(-\Delta_k) - n_F(\Delta_k)\bigr] \\
	&\qquad \qquad - \frac{k_x^2}{\Delta_k^3} 
\bigl[n_F(-\Delta_k) - n_F(\Delta_k) + \Delta_k \bigl(n_F'(\Delta_k) + 
n_F'(-\Delta_k)\bigr)\bigr]\Bigr\}.
\end{split}
\end{equation}
We are interested primarily in the limit $\omega \ll T$ in order to obtain the 
coefficient of $\delta(\omega)$. In this case, we can set $\omega + i \delta = 
0$ in the curly bracket, which yields
\begin{equation}
	= -\frac{i e_A^2 N}{\omega + i \delta} \intd{k}{2} 
\intd{K'}{\frac{1}{2}-\epsilon} 
\frac{k_x^2}{\Delta_k^2} \bigl[n_F'(\Delta_k) + n_F'(-\Delta_k)\bigr].
\end{equation}
Taking the real part, we obtain
\begin{equation}
\begin{split}
	\operatorname{Re} \sigma_{xx}^\text{1Loop}(\omega \ll T) &= -\pi 
\delta(\omega) e_A^2 N \intd{k}{2} \intd{K'}{\frac{1}{2}-\epsilon} 
\frac{k_x^2}{\Delta_k^2} 
\bigl[n_F'(\Delta_k) + n_F'(-\Delta_k)\bigr]\\
	& = 2 \pi e_A^2 N \delta(\omega) T^{1/2 - \epsilon} \intd{k_y}{} 
\frac{\pi^{3/4-\epsilon/2} (1 - 2^{1/2+\epsilon}) \Gamma(\frac{3}{2} - 
\epsilon) \zeta(\frac{1}{2} - \epsilon)}{(2\pi)^{3/2-\epsilon} 
\Gamma(\frac{7}{4} - \frac{\epsilon}{2})}.
\end{split}
\end{equation}
For $d = 2$ ($\epsilon = 1/2$), this result reduces to
\begin{equation}
	\operatorname{Re} \sigma_{xx}^\text{1Loop}(\omega) = e_A^2 N 
\delta(\omega) \intd{k_y}{},
\end{equation}
which coincides with the result that follows from Eq.~\eqref{eq:JJ_1Loop}.

\input{nematic_hyper3.bbl}

\end{document}

%% file: nematic_hyper3.bbl
%